\documentclass[draftcls,12pt,onecolumn]{IEEEtran}
\usepackage{amssymb}
\usepackage{cite}
\usepackage{url}
\usepackage{xcolor}
\usepackage{amsmath}
\usepackage{graphicx}
\usepackage{derivative}
\usepackage{algorithm}
\usepackage{algpseudocode}
\usepackage{mathtools}
\usepackage{balance}
\usepackage{commath}
\usepackage{epstopdf}
\usepackage{blkarray} %
\usepackage{adjustbox}
\usepackage{bm}
\usepackage[hidelinks]{hyperref}

\begin{document}

	\title{Energy-Efficient Aerial    Network Slicing for Computation Offloading, Data Gathering, and Content Delivery}
	\author{Ahmed A. Al-habob,~\IEEEmembership{Senior Member,~IEEE,} 
		Octavia A. Dobre,~\IEEEmembership{Fellow,~IEEE,}  and 
		Yindi Jing,~\IEEEmembership{Senior Member,~IEEE}
}



	\maketitle
	
\IEEEpeerreviewmaketitle
\begin{abstract}
	This paper introduces an unmanned aerial vehicle (UAV)-enabled network slicing problem to provide content delivery, sensing data gathering, and mobile edge computing (MEC) services. Three tenants provide services to their clients by sharing a common infrastructure of a set of UAVs. The content delivery tenant needs to guarantee that each of its clients (users) receives the required content,   the sensing tenant aims to gather an adequate amount of uncorrelated data, and the MEC tenant provides computing service to its clients.  An energy consumption minimization framework is considered to meet the tenants' requirements by optimizing the number of deployed UAVs, the deployment location of each UAV, the transmit power of each deployed UAV,   the user-UAV association, and the transmission power as well as the computing resources of each UAV. Taking into account the spatial correlation among the sensing users, a subset of these users is activated to gather the required sensing information. A solution approach technique inherited from graph theory is presented, in which the Lagrange approach derives the transmission power and computing resource allocation expressions. Simulation results illustrate that the proposed framework significantly reduces the total
	  energy consumption.
\end{abstract}
 \section{Introduction}
 The recent development of communication networks, such as the fifth generation (5G) and beyond 5G networks, is associated with a major development in the networks' infrastructure and resources \cite{10551400}.  
 Alongside the growth of communication networks, smartphone devices and vehicular systems exhibit more computing and sensing capabilities, which increases the amount and diversity of the gathered data.  The sensing capability of devices has given birth to an emerging paradigm, namely mobile crowd sensing (MCS), which enables devices to build participatory sensor networks \cite{9780171}. 
 The MCS devices can collect sufficient information regarding the detection of physical phenomena or surrounding conditions such as traffic or temperature. By designing appropriate rewards for the contributed devices, the participation of the devices can be decided.  The reward of each contributing device can be proportional to the importance or novelty of the data obtained by the device. Some work in the literature has investigated the aggregation of data from a group of devices considering the spatial correlation \cite{9217066,9409148} and/or the novelty of information \cite{9606181}.     
 
 Content delivery represents an important   aspect  of the 5G networks as daily generated content (e.g., YouTube videos, Dropbox shared files, Instagram pictures) witness exponential growth in their quality and quantity requirements. Multiple efforts have been devoted to developing different techniques for content delivery, such as network coding \cite{8169684} and unmanned aerial vehicle (UAV)-enabled data delivery scenarios \cite{9174931,9865119}.
  The evolution of 5G networks is associated with increased computing capabilities throughout these networks, especially at the edges, which gives birth to the mobile-edge computing (MEC) paradigm to provide task offloading services in terrestrial and non-terrestrial networks \cite{10963879}.  Integrating aerial devices (such as UAVs and high-altitude platforms) into 5G networks enables flexible and cost-efficient infrastructure deployment
   as well as providing services, including MEC services, in areas lacking ground infrastructure \cite{10547452,9598918,10339670}.  UAV-enabled  networks leverage the   line-of-sight (LoS) dominant UAV-ground communication channels, which  can  also include non-LoS
   	(NLoS) links in some locations with different characteristics, such as urban areas \cite{8698468}.

Network slicing enables the coexistence of multiple virtual networks that share the same
infrastructure while providing different services with heterogeneous requirements \cite{10242032}. Network slicing enables mobile network
operators to lease their communications resources, such as base stations, cell sites, and data
centers, to service providers or tenants
which offer services to their customers or users \cite{10679214}. The tenants  lease resource slices to meet their customers
 	demands while the network operators provide the resources \cite{10847822}. This concept is referred to as network-as-a-service (NaaS), which makes optimal allocation of the available communication infrastructure and
resources a critical challenge that must be addressed to meet the tenants' obligation and benefit the operators.

\begin{table*}[h!] 
	\caption{Main Notations Used in the Paper.}
	\begin{center}
		\begin{adjustbox}{width=.99\textwidth}
			\begin{tabular}{|c|c||c|c|}
				\hline
				{Notation} &  {Description} & {Notation} &  {Description}\\ \hline 
$  N $	&\multicolumn{1}{l||}{Total number of  users in the users set $  \mathcal{N}  $}  & $ U $ & \multicolumn{1}{l|}{Number of available UAVs in the UAVs set $\mathcal{U}$}           \\ \hline   
$ \mathcal{N}_c/ \mathcal{N}_s/ \mathcal{N}_m $	&\multicolumn{1}{l||}{Set of content delivery/sensing/MEC users, $\mathcal{N}=\{\mathcal{N}_c, \mathcal{N}_s, \mathcal{N}_m \} $}      & $ 	K$  &\multicolumn{1}{l|}{Number of deployed UAVs  $K \leq U$}         \\ \hline
$ N_c/N_s/N_m $	&\multicolumn{1}{l||}{Number of content delivery/sensing/MEC users, {$N=N_c+N_s+N_m$}}      & {$ 	C$}  &\multicolumn{1}{l|}{Number of contents   in the content catalog   $ 	\mathcal{C}$  }         \\ \hline
$n_i/u_k/c_j  $	&\multicolumn{1}{l||}{ User $n_i$/ UAV $u_k$/ content item $c_j$}      & $ M_j$  &\multicolumn{1}{l|}{Size of content $c_j$ (in bits)}         \\ \hline
				$ \mathbf{R}=\left[r_{ij} \right]_{\footnotesize{ N_c\times C}} $ 	&\multicolumn{1}{l||}{Content demand indicator, $r_{ij}$ is defined in \eqref{demandind}    }         & $ 	\mathbf{S}=\left[s_{jk} \right]_{\footnotesize{ C\times U}} $ &\multicolumn{1}{l|}{UAVs' content storage indicator, $s_{jk}$ is defined in \eqref{StorInd} }      \\ \hline
				$L$  	&\multicolumn{1}{l||}{Size of the raw data gathered by a sensing user}      & $H$     &\multicolumn{1}{l|}{Size of the uncorrelated data (information) of all sensing users }        \\ \hline
				$ \bm{\eta}=\left[\eta_{i} \right]_{\footnotesize{ N_s\times 1}} $		&\multicolumn{1}{l||}{Sensing users activation decision variable, $\eta_{i}$ is defined in \eqref{sensActiv}  }            &$ \mathcal{H}(\bm{\eta})$    & \multicolumn{1}{l|}{Size of  information of the active sensing users}             \\ \hline
				$ \rho $ 		
				&\multicolumn{1}{l||}{Data correlation extent parameter}               & $\ell_i/ \varsigma_i$     &\multicolumn{1}{l|}{Task size/number of CPU cycles  to compute one bit of user $n_i \in \mathcal{N}_m$}   \\ \hline
				$	\mathbf{h}_{ik}$ 	&\multicolumn{1}{l||}{Communication channel vector between a user $n_i$ and a UAV $u_k$ }            &  $\psi_k=\{x_k, y_k, z_k\}$     &\multicolumn{1}{l|}{Cartesian coordinates of the placement location of $u_k$}              \\ \hline
				$ d_{ik}=\| \psi_k-  \bar{\psi}_i\| $   		
				&\multicolumn{1}{l||}{Distance between the user $ n_i $ and   UAV  $ u_k $  }              &$\bar{\psi}_i=\{\bar{x}_i, \bar{y}_i, 0\}$     &\multicolumn{1}{l|}{Cartesian coordinates of user $n_i$}                 \\ \hline
				$ \lambda_0 $   	 &\multicolumn{1}{l||}{Path loss at the reference distance of 1 meter }             & $ \beta/F $   &\multicolumn{1}{l|}{Path loss exponent/Rician factor }             
				\\ \hline
				$P_k/F_k/A_k$ 	 &\multicolumn{1}{l||}{Transmit power/computing speed/number of antennas of UAV $u_k$  }             & $\tilde{\mathbf{h}}^{\mbox{\scriptsize LoS}}_{ik}/\tilde{\mathbf{h}}^{\mbox{\scriptsize NLoS}}_{ik}$      &\multicolumn{1}{l|}{LoS/NLoS components of the channel $	\mathbf{h}_{ik}$ }             \\ \hline
				$\phi_{ik}$	 &\multicolumn{1}{l||}{Angle between user $n_i$ and UAV $u_k$}             &  $ \bm{\mu}=\left[\mu_{ik} \right]_{\footnotesize{ N\times U}} $    &\multicolumn{1}{l|}{User-UAV association decision variable, $\mu_{ik}$ is defined in \eqref{USerUAV} }             
				\\ \hline
				$P_{ik}$	 &\multicolumn{1}{l||}{The allocated transmit power  for user $n_i$ at UAV $u_k$ }             & $ \mathbf{w}_{ik} T$ &\multicolumn{1}{l|}{Beamformer for user $n_i$ at UAV $u_k$}             \\ \hline
				$\textbf{P}\triangleq[\textbf{p}_1, \cdots, \textbf{p}_K]$   	 &\multicolumn{1}{l||}{Power allocation decision with $\textbf{p}_k=[P_{1k}, \cdots, P_{Nk}]^T$}             &  		 
				${\gamma}_{ik}(\bm{\eta},\bm{\mu},\bm{\psi})$ &\multicolumn{1}{l|}{  Signal-to-interference-plus-noise ratio at user $n_i$  from UAV $u_k$}                      \\ \hline
				{	${\mathcal{R}}_{ik}(\bm{\eta},\bm{\mu},\bm{\psi})$	} &\multicolumn{1}{l||}{Data rate at user $n_i$ from UAV $u_k$}             &
				$\bar{\gamma}_{ik}(\bm{\eta},\bm{\mu},\bm{\psi})$ &\multicolumn{1}{l|}{  Signal-to-interference-plus-noise ratio at UAV $u_k$ from user $n_i$ }                      \\ \hline
				{	$\bar{\mathcal{R}}_{ik}(\bm{\eta},\bm{\mu},\bm{\psi})$	} &\multicolumn{1}{l||}{Data rate at UAV $u_k$ from user $n_i$}        &  $ V $  &\multicolumn{1}{l|}{Travelling speed of  the UAV}                          \\ \hline
				{$ P^{(\mbox{\scriptsize prof})}/P^{(\mbox{\scriptsize ind})}  $	} &\multicolumn{1}{l||}{Blade profile/induced power of the UAV in
					hovering status }             & $ \varpi $ &\multicolumn{1}{l|}{Rotor blade's  tip speed}                          \\ \hline
				{$ v_0 $	} &\multicolumn{1}{l||}{Hovering mean rotor-induced velocity}           & 
				$ \delta_0/\zeta $ &\multicolumn{1}{l|}{Fuselage drag ratio/rotor solidity}                          \\ \hline
				{$ \varsigma/\xi $	} &\multicolumn{1}{l||}{Air density/rotor
					disc area}             & 
				$ 	P^{(\mbox{\scriptsize hov})} $ &\multicolumn{1}{l|}{Hovering power of the UAV}                          \\ \hline
				{$ \psi_k^0 $	} &\multicolumn{1}{l||}{Docking location of UAV $u_k$}             & 
				$ \mathcal{E}^{\mbox{\scriptsize mov}}_k(\bm{\psi})  $ &\multicolumn{1}{l|}{ Energy consumption of moving UAV $u_k$ from  $ \psi_k^0 $ to $ \psi_k $}                          \\ \hline				
				{$ 	\mathcal{E}^{\mbox{\scriptsize hov}}_k(\bm{\eta},\bm{\mu},\bm{\psi}, \textbf{f},\textbf{P}) $	} &\multicolumn{1}{l||}{Energy consumption of   UAV $u_k$ while hovering at $ \psi_k $}             & 
				$ \mathcal{T}_{ik}(\bm{\mu},\textbf{f})$ &\multicolumn{1}{l|}{Task offloading latency from user $n_i$ to UAV $u_k$}                          \\ \hline				
				{$  \mathbf{f}_k\triangleq[f_{1k}, \cdots, f_{N_mk}] $	} &\multicolumn{1}{l||}{Computing resource allocation decision at   $ u_k $   }             & 
				$ f_{ij} $ &\multicolumn{1}{l|}{Computational speed allocated to  $n_i$ at   $ u_k$}                          \\ \hline				
				{$\mathcal{E}^{\mbox{\scriptsize comp}}_k(\bm{\mu},\textbf{f})$} &\multicolumn{1}{l||}{Computing energy consumption of the CPU of UAV $ u_k $ }             & 
				$\mathcal{E}^{\mbox{\scriptsize tra}}_k(\bm{\mu}, \bm{\psi},\textbf{P})$  &\multicolumn{1}{l|}{Transmission energy consumption of UAV $u_k$}                          \\ \hline				
				{$	\mathcal{E}_k(\bm{\eta},\bm{\mu},\bm{\psi},\! \textbf{f},\!\textbf{P})$  	} &\multicolumn{1}{l||}{Total energy consumption of   UAV $ u_k $}             & 
				$ z^{\min}/z^{\max}$  &\multicolumn{1}{l|}{Minimum/maximum allowable altitude limits of UAV $u_k$}                          \\ \hline		  						
				{$\mathcal{G}\triangleq\{\mathcal{V}, \mathcal{E}, \mathcal{W}\}$} &\multicolumn{1}{l||}{Graph model with    $\mathcal{V}$  vertices,  $ \mathcal{E}$  edges, and $\mathcal{W}$   weight of the edges}             & 
				$ \mathcal{V}^c/\mathcal{V}^s/\mathcal{V}^m $ &\multicolumn{1}{l|}{Vertices sets of content/sensing/ MEC users}                          \\ \hline
				{$w_{ik}(\hat{\iota})$} &\multicolumn{1}{l||}{The weight of the edge $i,k$ at the $\hat{\iota}$-th step}             & 
				$\mathcal{V}^u $ &\multicolumn{1}{l|}{Vertices sets of the UAVs}                          \\ \hline
				$Q$ &\multicolumn{1}{l||}{Number of iterations in Algorithm $3$}             & 
				$I $ &\multicolumn{1}{l|}{The amount of the required information by the sensing tenant }                          \\ \hline				
			\end{tabular}
		\end{adjustbox}
	\end{center}
	\label{mysymbols}
\end{table*}

Various works in the literature have considered UAV-enabled MEC and network slicing.   In \cite{10988691},     
a collaborative multi-UAV  decision-making system  was considered to optimize the task offloading and resource allocation in MEC networks. The objective was
to reduce the task computing delay and energy consumption under constraints on   allowable task completion time and compliance with resource limitations. A  two-stage optimization algorithm was developed  to   optimize the task offloading decision and resource allocation of the collaborative computing system. In \cite{11077695},  
a hierarchical aerial MEC network architecture was considered to improve the quality of user experiences. A bi-level UAV-enabled MEC network   was   deployed to provide continuous MEC services to ground users with variable demands. An optimization problem was formulated to minimize the utility of all users. The stability of task queue backlogs and energy consumption budgets at users and UAVs were considered. In \cite{11075956},  
a multi-domain network slicing scheme for satellite-airborne-terrestrial edge computing networks was considered. Each slice has configured to include terrestrial-airborne, terrestrial-satellite, or terrestrial-airborne-satellite domain typologies based on the  resources availability. The paper optimized the slice configuration selection, routing, and resource allocation.   
	In \cite{9127468}, a UAV network slicing framework was considered in which 
	a system controller   can turn on and off the computing elements at the UAVs, with the possibility of offloading jobs to other UAVs.
	In \cite{10679214}, a hierarchical UAV slicing framework that operates at two different time-scales was studied. The problem of inter-slice resource management was formulated as a mixed integer nonlinear program and a stochastic game with the objective  of maximizing the total transmission rate. 
The aforementioned works did not consider providing multiple services using a shared UAV networks, with each service has its own requirements.

This paper develops a framework to enable a content dissemination tenant, a sensing data gathering tenant, and an MEC tenant
to share a set of UAVs to provide services to their users. The content dissemination tenant aims to deliver the required content to each of its subscribers or users. The objective of the sensing data gathering tenant is to gather a sufficient amount of uncorrelated data from his users while the MEC tenant provides computing services. 
The developed framework enables the software-defined networking (SDN)/ network function virtualization (NFV) controller to satisfy the tenants' requirements with minimum energy consumption. As the network conditions and/or the tenants' requirements change, the developed framework enables the SDN/NFV to decide the number of deployed UAVs and their deployment locations,  allocate the transmitted power and the computation resources of the deployed UAVs,  select the active sensing devices, and obtain the user-UAV association.
The main contributions of this paper can be summarized as follows:

\begin{itemize}
	\item[$\bullet$] A network slicing framework is developed to share a multi-UAV network by content dissemination, sensing data gathering, and MEC tenants.
	\item An energy minimization problem is formulated to guarantee that all the required contents are delivered,   sufficient information is gathered, and all the MEC users are served. The UAVs' deployment locations, the UAVs' power allocation, computing resources allocation, and the user-UAV association are optimized.  Considering the spatial correlation among the users and the trade-off between the consumed energy and the number of active sensing users,    the sensing users' activation is also optimized. 
	\item[$\bullet$] A  solution inherited from graph theory is devised, in which the Lagrange approach derives the transmission   
	power and computing resource allocation expressions.
\end{itemize}

The remainder of this paper is organized as follows.
Section \ref{Sys}  presents the considered system model and discusses  data gathering, communication, and energy consumption models.      In Section \ref{ProbFormulation}, the energy consumption minimization problem is formulated. Section \ref{Sol} introduces the proposed heuristic solution.     
Section \ref{SR} discusses the   simulation
results, and Section \ref{CO} concludes the paper.

	\textit{Notation:} Lower  case and upper case boldface letters  denote vectors and matrices, respectively. $(\cdot)^H$ and	$(\cdot)^T$    represent the Hermitian and the transpose     operators, respectively. The main notations used throughout this paper are summarized in Table \ref{mysymbols}.

\begin{figure}[h!]
	\centering
	\includegraphics[width=0.6\textwidth]{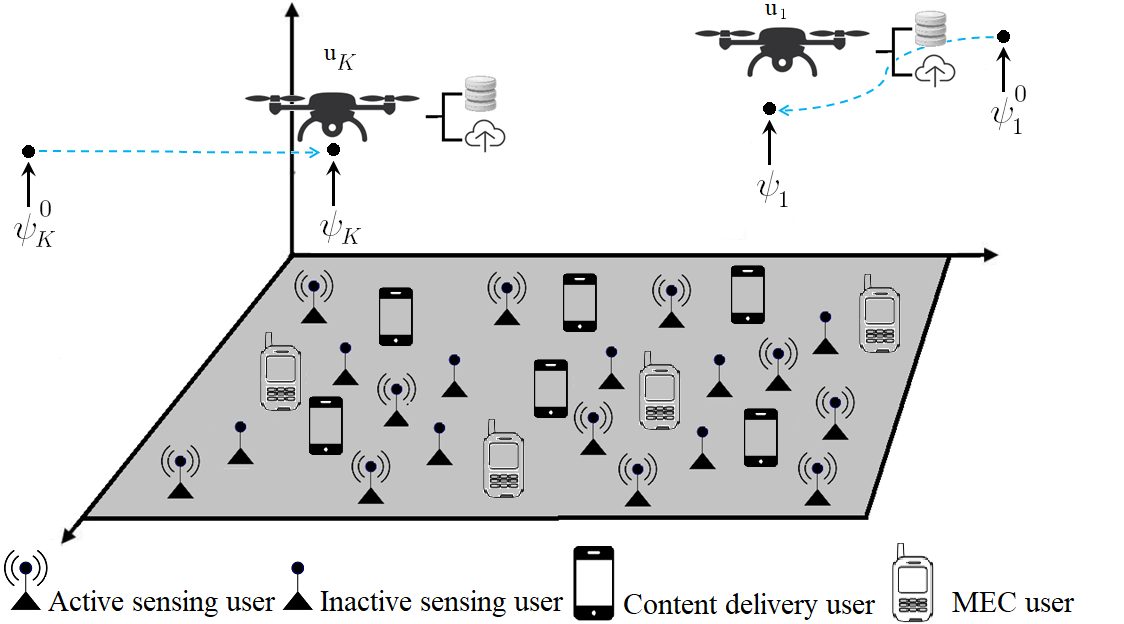}
	\caption{System model with   $K\leq U$ deployed UAVs.}
	\label{Sys_Model2}
\end{figure}


\section{System Model}\label{Sys}
As shown in Fig. \ref{Sys_Model2}, a network slicing framework is considered where three
tenants (i.e., services) provide content delivery,   sensing data gathering, and MEC   services to a set of $\mathcal{N} = \{n_i\}_{i=1}^{N}$ of $N$ users.  The UAV-enabled network consists of a set     $\mathcal{U} = \{u_k\}_{k=1}^{U}$ of $U$  UAVs; each UAV operates in full duplex mode with a maximum transmit power $P_k$, computing speed $F_k$,  and  $A_k$ antennas. The tenants share $K$ UAVs, where $K \leq U$ is the number of deployed UAVs.
 Each tenant  serves its own subset of users $\mathcal{N}=\{\mathcal{N}_c, \mathcal{N}_s, \mathcal{N}_m \}$, where  $\mathcal{N}_c = \{n_{i}\}_{i=1}^{N_c}$ represents the content   delivery users,   $\mathcal{N}_s = \{n_{i}\}_{i=N_c+1}^{N_c+N_s}$ is the set of sensing users, and $\mathcal{N}_m = \{n_{i}\}_{i=N_c+N_s+1}^{N}$ is the MEC users where $N=N_c+N_s+N_m$.
The content delivery tenant serves its users by delivering a catalog     $\mathcal{C} = \{c_j\}_{j=1}^{C}$ of $C$ contents each of size $M_j$ bits. A content delivery user is interested in downloading a content. To represent the users' content demand, let us define $ 	\mathbf{R}=\left[r_{ij} \right]_{\footnotesize{ N_c\times C}} $ such that    
\begin{equation}\label{demandind}
r_{ij} = \begin{cases} 1, &\mbox{if} ~n_{i}~  \mbox{is interested in downloading}~ c_j,   \\ 
		0, & \mbox{otherwise}. \end{cases}
\end{equation} 
Furthermore, let us define the UAVs' content storage indicator $ 	\mathbf{S}=\left[s_{jk} \right]_{\footnotesize{ C\times U}} $ such that    
\begin{equation}\label{StorInd}
	s_{jk} = \begin{cases} 1, &\mbox{if} ~c_j~  \mbox{is stored in}~ u_k,   \\ 
		0, & \mbox{otherwise}. \end{cases}
\end{equation} 
Additionally,  $\sum_{j=1}^{C}s_{jk}\geq1, \forall c_j \in \mathcal{C} $ to reflect the fact that each content should be stored in at least one UAV.

The sensing tenant gathers data from  its users, where each user can obtain  $L$
bits of raw data. Activating all the users is an inefficient approach as their data could be correlated; this approach yields more interference at the UAVs and increases their travelling and hovering time. The spatial correlation, which is a result of the close proximity between the sensing users, is the main cause of data correlation.  Consequently, the sensing tenant aims at gathering uncorrelated data (referred to as information in this paper).
 An empirical study has been performed in \cite{pattem2008impact}  to quantify the total gathered information from a set of $N_s$ sensing users using a
constructive iterative approach. Since a sensing user generates $L $ (bits) of raw data, the information provided by the first sensing user is $\mathcal{H}_1= L $ (bits). The information collected by the first and second active sensing users can be obtained as $ \mathcal{H}_2= L + \left[1-\frac{1}{\left(\bar{d}_{1,2}/\rho +1\right)}\right]L $, where  $ \bar{d}_{1,2} $ is the Euclidean  distance between the locations of the first and second active sensing users and
  $ \rho $ is a  parameter that depends on  the  
correlation extent in the sensed data \cite{pattem2008impact}. The information
of the first three active sensing users can be obtained as
\begin{equation}
	\mathcal{H}_3= L + \left[1-\frac{1}{\left(\bar{\bar{d}}_2/\rho +1\right)}\right]L +  \left[1-\frac{1}{\left(\bar{\bar{d}}_3/\rho +1\right)}\right]L,
\end{equation}
where $ \bar{\bar{d}}_2 =\bar{d}_{1,2} $ and $ \bar{\bar{d}}_3 =\min\{\bar{d}_{1,3}, \bar{d}_{2,3} \} $.  Consequently,      the  maximum information that can be gathered by $ N_s$ sensing users can be expressed as
\begin{equation}\label{H}
	{H}= L+L\sum\limits_{i=2}^{N_s}      \left[1-\frac{1}{\bar{\bar{d}}_i/\rho +1}\right],
\end{equation}
where $\bar{\bar{d}}_i$ is the minimum distance between the location of the sensing user    $ n_i $ and all other previously considered users  $ n_\iota $   $\forall \iota=1, 2, \dots, i-1$.
Let us define the sensing users activation decision $ \bm{\eta}=\left[\eta_{i} \right]_{\footnotesize{ N_s\times 1}} $ such that    
\begin{equation}  \label{sensActiv}
	\eta_{i} = \begin{cases} 1, &\mbox{if} ~n_{i}~  \mbox{is active},    \\ 
		0, & \mbox{otherwise}. \end{cases}
\end{equation} 
For a given  sensing users activation decision $\bm{\eta}$, the gathered information of the active sensing users (users with $\eta_{i}=1$) can be obtained as 
\begin{equation}\label{Heta}
	\mathcal{H}\left(\bm{\eta}\right)=   L+ L\sum\limits_{i=2}^{N_s} \eta_{i}    \left[1-\frac{1}{ {d}_i\left(\bm{\eta}\right)/\rho +1}\right] ,  
\end{equation} 
where   ${d}_i\left(\bm{\eta}\right)$ is the minimum distance between the location of active user $ n_i $ and all  active users  with $ \eta_{\iota}=1 $ $\forall \iota=1, 2, \dots, i-1$.

 Each MEC user offloads a task $\{\ell_i, \varsigma_i\}$ $\forall ~ n_i \in \mathcal{N}_m$, where  $\ell_i $ is the task size (in bits) and $\varsigma_i$ is the number of CPU cycles required to compute one bit. 
 
\subsection{Communication Model}
The communication channel between a user $n_i$ and a UAV $u_k$   is modelled as
\begin{equation}\label{channel}
	\mathbf{h}_{ik}=\sqrt{\lambda_0 d_{ik}^{-\beta}}\left[\sqrt{\frac{F}{F+1}}\tilde{\mathbf{h}}^{\mbox{\scriptsize LoS}}_{ik}
	+\sqrt{\frac{1}{F+1}}\tilde{\mathbf{h}}^{\mbox{\scriptsize NLoS}}_{ik}\right],
\end{equation}
where $ d_{ik}=\| \psi_k-  \bar{\psi}_i\| $ is the distance between the user $ n_i $ and   UAV  $ u_k $, where 
$\psi_k=\{x_k, y_k, z_k\}$ are the Cartesian coordinates of the placement location of $u_k$ and  $\bar{\psi}_i=\{\bar{x}_i, \bar{y}_i, 0\}$ represent the Cartesian coordinates of user $n_i$.
  $ \lambda_0 $ is the path loss at the reference distance of 1 meter, $ \beta $ is the path loss exponent, $ F $   denotes the
Rician factor, $ \tilde{\mathbf{h}}^{\mbox{\scriptsize NLoS}}_{ik} \in \mathbb{C}^{ A_k\times 1} $ represents the non-line-of-sight scattering components modelled by independent random Rayleigh
distributed entries, and $ \tilde{\mathbf{h}}^{\mbox{\scriptsize LoS}}_{ik} \in \mathbb{C}^{ A_k\times 1} $ is the   LoS component. By considering a uniform linear array with the inter-element spacing being one-half the carrier wavelength, the LoS component of the channel from the device $n_i$ to the UAV-mounted server $u_k$ is modelled as
\begin{equation}
	\tilde{\mathbf{h}}^{\mbox{\scriptsize LoS}}_{ik}\!\!=[1, e^{-\sqrt{-1}\pi\cos \phi_{ik}}, \cdots, e^{-\sqrt{-1}{\pi(A_k-1)}\cos \phi_{ik}}],
\end{equation}
where     $ \phi_{ik} $ is the   angle between   user $ n_i $ and the UAV $ u_k $, such that $ \cos \phi_{ik}= \frac{x_k-\bar{x}_i}{\|\psi_k-\bar{\psi}_i\|} $.
Each UAV operates on its orthogonal channel and is deployed to serve a subset of users $\mathcal{N}_k \subseteq \mathcal{N}$, such that $\bigcup\limits_{k=1}^{K}\mathcal{N}_k=\mathcal{N}$. 
Let us define the user-UAV association decision $ \bm{\mu}=\left[\mu_{ik} \right]_{\footnotesize{ N\times U}} $ such that    
\begin{equation}\label{USerUAV}
\mu_{ik} = \begin{cases} 1, &\mbox{if} ~n_{i}~  \mbox{is served by } u_k,   \\ 
		0, & \mbox{otherwise}. \end{cases}
\end{equation}
Let $\textbf{P}\triangleq[\textbf{p}_1, \cdots, \textbf{p}_K]$ with $\textbf{p}_k=[P_{1k}, \cdots, P_{Nk}]^T$  as the UAV  transmit power allocation decision of UAV $u_k$,  the signal-to-interference-plus-noise ratio
(SINR) at user $n_i$  to decode the message sent by $u_k$  can be expressed as 
\begin{equation}
	\gamma_{ik}(\bm{\mu},\bm{\psi}, \textbf{P})=  \frac{\mu_{ik}P_{ik}|{\mathbf{h}^H_{ik}}{\mathbf{w}_{ik}}|^2}{\sum\limits_{\substack{i'=1\\i'\ne i}}^{N}\mu_{i'k}P_{i'k}|{\mathbf{h}^H_{ik}}{\mathbf{w}_{i'k}}|^2
		+ \sigma^2_i}, 
\end{equation}
where $ P_{ik}$ is the allocated power by the UAV $u_k$ to communicate with user $n_i$ and $\mathbf{w}_{ik}$ is the beamforming vector, which can be obtained using the maximum ratio transmission  
scheme    as follows
\begin{equation}\label{beam}
	\mathbf{w}_{ik} =	\frac{ \mathbf{{h}}_{ik} }{\|\mathbf{{h}}_{ik}\|}.
\end{equation}
The corresponding data rate is 
\begin{equation}\label{ratte}
	\mathcal{R}_{ik}(\bm{\mu},\bm{\psi}, \textbf{P}) =B_k \log_2 \left(1+\gamma_{ik}(\bm{\mu},\bm{\psi}, \textbf{P})\right),
\end{equation}
where $B_k$ is the downlink bandwidth of UAV $u_k$. The SINR  of user $n_i$ at UAV $u_k$ can be expressed as
\begin{equation}
	\bar{\gamma}_{ik}(\bm{\eta},\bm{\mu},\bm{\psi})=  \frac{\eta_i\mu_{ik}\bar{P}_{i}|{\mathbf{h}^H_{ik}}{\mathbf{\bar{w}}_{ik}}|^2}{\sum\limits_{\substack{\iota=1\\ \iota\neq i}}^{N}\eta_{\iota} \mu_{\iota k}  \bar{P}_{\iota}|{\mathbf{h}^H_{ik}}{\mathbf{\bar{w}}_{\iota k}}|^2+ \sigma^2_k}, 
\end{equation} 
where $ \bar{P}_{i}$ is the transmitted power of user $n_i$ and $\mathbf{\bar{w}}_{ik}$ is the beamforming vector, which can be obtained using the maximum ratio combining  
scheme.  The uplink bandwidth is $ \bar{B}_k$, and thus
the   data rate from $n_i$ to $u_k$ is expressed as
\begin{equation}\label{ratte_bar}
	\bar{\mathcal{R}}_{ik}(\bm{\eta},\bm{\mu},\bm{\psi}) =\bar{B}_k \log_2 \left(1+\bar{\gamma}_{ik}(\bm{\eta},\bm{\mu},\bm{\psi})\right).
\end{equation}

	\subsection{Energy Consumption Model}

The UAV   consumes its energy to realize the following primary functions: travelling, hovering, and communication. According to \cite{8663615},  the movement
power consumption of a UAV can be modelled as

\begin{equation}\label{Trpower}
	\begin{split}
		P^{(\mbox{\scriptsize mov})}=& P^{(\mbox{\scriptsize prof})}\left[1+\frac{3V^2}{\varpi^2}\right] +P^{(\mbox{\scriptsize ind})}\left[\sqrt{1+\frac{V^4}{4v^4_0}}-\frac{V^2}{2v_0^2}\right]^{\frac{1}{2}} +  \frac{1}{2}\delta_0\varsigma \zeta\xi V^3,
	\end{split}
\end{equation}
where $ V $ is the travelling speed of  the UAV, $ P^{(\mbox{\scriptsize prof})} $ and $P^{(\mbox{\scriptsize ind})}  $ are the power of the blade profile  and induced power  in
hovering status, respectively. The rotor blade's  tip speed is represented by  $ \varpi $, $ v_0 $ is   the hovering mean rotor-induced velocity, $ \delta_0 $ and $ \zeta $ denote the fuselage drag ratio and rotor solidity,
respectively. Finally,    $ \varsigma $ and $ \xi $ represent the air density and rotor
disc area, respectively. 
It is worth mentioning that the power consumed by the UAV in hovering $ 	P^{(\mbox{\scriptsize hov})} $ can be obtained by setting $ V=0 $ in \eqref{Trpower}, which leads to $ 	P^{(\mbox{\scriptsize hov})} = P^{(\mbox{\scriptsize prof})} + P^{(\mbox{\scriptsize ind})}$    \cite{8663615}.  Consequently, the energy consumption of flying  UAV $u_k$ from its docking location $ \psi_k^0 $ to     the placement location $ \psi_k $ is modeled as \cite{8663615,8613833}

\begin{equation}\label{E_tra}
	\begin{split}
		\mathcal{E}^{\mbox{\scriptsize mov}}_k(\bm{\psi}) =& \frac{P^{(\mbox{\scriptsize mov})}}{V} \|\psi_k^0-\psi_k\|.
	\end{split}
\end{equation}

The energy consumption of   the UAV $ u_k$  while hovering at its placement location can be expressed as 
\begin{equation}\label{E_hov}
	\begin{split}
	\mathcal{E}^{\mbox{\scriptsize hov}}_k(\bm{\eta},\bm{\mu},\bm{\psi}, \textbf{f},\textbf{P})=& P^{(\mbox{\scriptsize hov})}
		\max\limits_{\substack{\forall n_i\in \mathcal{N}}}\Biggl\{  \frac{\mu_{ik}\sum\limits_{j=1}^{C}r_{ij}M_j}{\mathcal{R}_{ik}(\bm{\mu},\bm{\psi}, \textbf{P})},  \frac{\eta_{i}\mu_{ik}L}{\bar{\mathcal{R}}_{ik}(\bm{\eta},\bm{\mu},\bm{\psi})},   ~\mathcal{T}_{ik}(\bm{\mu},\textbf{f})\Biggr\},
	\end{split}
\end{equation}     
where $ \mathcal{T}_{ik}(\bm{\mu},\textbf{f})$ is the offloading latency which can be expressed as 
\begin{equation}
	\mathcal{T}_{ik}(\bm{\mu},\textbf{f}) = \begin{cases} \mu_{ik}\left(\frac{\ell_i}{\bar{\mathcal{R}}_{ik}} + \frac{\ell_i\varsigma_i}{f_{ik}}\right), & \forall n_i \in \mathcal{N}_m,   \\ 
		0, & \mbox{otherwise}, \end{cases}
\end{equation}
where   $ f_{ij} $ is the computational speed allocated to  $n_i$ at   $ u_k$ (cycle per second). The   computing resource allocation decision at   $ u_k $ is   $  \mathbf{f}_k\triangleq[f_{1k}, \cdots, f_{N_mk}] $, which  satisfies  $\sum_{i=1}^{N_m}f_{ik} \leq F_k$, where $F_k$ is the total computing resource of   $ u_k $.
The computing power consumption can be modeled as $ P^{(\mbox{\scriptsize comp})} = \kappa f^3$,   where $ f $ is the CPU’s computational speed  and    $ \kappa  $ is the effective switched capacitance depending on
	the CPU chip architecture \cite{7542156}.	Consequently, the computing energy consumption of the CPU of  $ u_k $   is expressed as 
	\begin{equation}\label{E_comp}
		\begin{split}
			\mathcal{E}^{\mbox{\scriptsize comp}}_k(\bm{\mu},\textbf{f})=& \kappa \sum_{i=1}^{N_m}\ell_i\varsigma_i f_{ik}^2 \mu_{\tilde{\tilde{i}}k},
		\end{split}
	\end{equation}
where $\tilde{\tilde{i}}=i+N_c+N_s$. The total energy consumed by UAV $u_k$ can be expressed as 
\begin{equation}\label{E_k}
	\begin{split}
	\!\!	\mathcal{E}_k(\bm{\eta},\bm{\mu},\bm{\psi},\! \textbf{f},\!\textbf{P})\!=&
	\mathcal{E}^{\mbox{\scriptsize mov}}_k\!(\bm{\psi})\!+\!
	 \mathcal{E}^{\mbox{\scriptsize hov}}_k\!(\bm{\eta},\bm{\mu},\!\textbf{f},\!\textbf{P})\! + \!	\mathcal{E}^{\mbox{\scriptsize comp}}_k(\bm{\mu},\!\textbf{f})+ \underbrace{\sum_{i=1}^{N_c} \frac{P_{ik} \mu_{ik}\sum\limits_{j=1}^{C}r_{ij}M_j}{\mathcal{R}_{ik}(\bm{\mu}, \bm{\psi}, \textbf{P})}}_{\mathcal{E}^{\mbox{\scriptsize tra}}_k(\bm{\mu}, \bm{\psi}, \textbf{P})},
	\end{split}
\end{equation}
where $\mathcal{E}^{\mbox{\scriptsize tra}}_k(\bm{\mu}, \bm{\psi},\textbf{P})$ is the transmission energy  consumed by the transmitter circuitry of  UAV $u_k$.  The energy consumed by the UAV's receiver circuitry is very small, and it can be neglected.

\section{Problem Formulation} \label{ProbFormulation}
 Let  $ I \leq H $ (in bits) be the minimum amount of information required to be aggregated to meet the sensing tenant requirement. The decision-makers need to satisfy the requirements of the tenants (aggregate the required information,   deliver all the required contents,  and compute the offloaded tasks). The objective is to minimize the energy consumption by deciding  the number of  deployed UAVs $K\leq U$ and their deployment locations   $\bm{\psi}$, selecting  the active sensing devices $\bm{\eta}$, allocating the transmitted power $\textbf{P}$ and the computation resources $\textbf{f}$, and determining the user-UAV association $\bm{\mu}$. The optimization problem is formulated as follows:

\begin{subequations}\label{P1}
	\begin{alignat}{2}
		\mbox{\textbf{P1}}~	&\!\min_{\substack{\bm{\eta},\bm{\mu}, \bm{\psi}\\\textbf{P}, \textbf{f}, K}}&&\sum_{k=1}^{K}\mathcal{E}_k\left(\bm{\eta},\bm{\mu},\bm{\psi},\textbf{f},\textbf{P}\right), \label{OPP1}\\
		&\mbox{s.t.}&&\sum_{k=1}^{U}\mu_{ik} \geq  1, ~  1 \leq i\leq N_c,
		\label{con1}\\	
		&&~~   &   \sum_{k=1}^{U} \sum_{j=1}^{C}\mu_{ik}s_{jk} r_{ij} =  \sum_{j=1}^{C}r_{ij},  ~  1 \leq i\leq N_c,
		\label{con22}\\
		&&~~   &    \sum_{k=1}^{U}\mu_{ik} =  \eta_i,  ~   N_c+1 \leq i\leq N_c+N_s ,
		\label{con2}\\
	    &&~~   &    \mathcal{H}\left(\bm{\eta}\right)\geq I, \label{con3}\\
	    &&~~   & \sum_{i=1}^{N_c} \mu_{ik}P_{ik}  \leq P_k, \forall u_k \in \mathcal{U}, \label{con4}\\
	    &&~~   &   \sum_{k=1}^{U}\mu_{ik} =  1, ~   N_c+N_s \leq i\leq N,
	    \label{con222}\\
	    &&~~   & \sum_{i=1}^{N_m} \mu_{\tilde{\tilde{i}}k} f_{ik}  \leq F_k, \forall u_k \in \mathcal{U}, \label{con5}\\
	    &&~~   & z^{\min} \leq z_k \leq z^{\max}, 1 \leq k\leq K, \label{con6}\\
		&&     & K \leq U, \mu_{ik}\mbox{ and }\eta_{i}\in \{0, 1\},  ~ \forall   n_i\in \mathcal{N}, u_k\in \mathcal{U}. \label{con7}
	\end{alignat}
\end{subequations}
Constraints \eqref{con1} and \eqref{con22} guarantee that each content delivery user is served by  one or more UAVs   that   store  the requested contents, respectively. Constraint \eqref{con2}  guarantees that each active sensing user is served by one UAV.
  Constraint \eqref{con3} guarantees that sufficient information is gathered.  Constraint \eqref{con4} represents the maximum transit power limit of each UAV. Constraints \eqref{con222} and \eqref{con5} guarantee that each MEC user is associated with a UAV and each UAV allocates the available computation resources, respectively. Constraint \eqref{con6} guarantees that each deployed UAV hovers within the allowable altitude limits.

\section{Proposed  Solution Approach}\label{Sol}
In this section, a  heuristic solution is developed to solve the formulated optimization problem  using techniques inherited from the
graph theory and the Lagrange approach. The proposed algorithm is a problem-specific approach that is designed based on the structure of the optimization problem, and it provides near-optimum energy consumption minimization in the considered system model. The proposed algorithm optimizes multiple decision variables including the number of deployed UAVs $K\leq U$, their deployment locations   $\bm{\psi}$ and the active sensing devices $\bm{\eta}$; allocates the transmitted power $\textbf{P}$ and the computation resources $\textbf{f}$; and determines the user-UAV association $\bm{\mu}$. The computation complexity of the algorithm is discussed in Section IV.D.

	\subsection{Optimizing the Transmission Power Allocation}
	
	For a given  $\bm{\eta},\bm{\mu}, \bm{\psi},   \textbf{f}$, and $K$,  the transmit power allocation 
	of UAV $u_k$ can be obtain by solving the following optimization problem
	\begin{subequations}\label{P1_P}
		\begin{alignat}{2}
			\mbox{\textbf{P1-P}}~	&\!\min_{\substack{\textbf{P}_k}}&&\max\limits_{\substack{\forall n_i\in \mathcal{N}_c}}\!\!\left\{\!  P^{(\mbox{\scriptsize hov})}  \mu_{ik} \frac{\sum\limits_{j=1}^{C}r_{ij}M_j}{\mathcal{R}_{ik}}, E_2 \right\}+\sum_{i=1}^{N_c}P_{ik} \mu_{ik} \frac{\sum\limits_{j=1}^{C}r_{ij}M_j}{\mathcal{R}_{ik}}, \\
			&\mbox{s.t.}&& \sum_{i=1}^{N_c} \mu_{ik}P_{ik}  \leq P_k,   \label{con1P}\\
			&&     &  P_{ik} \geq 0 ~ \forall   n_i\in \mathcal{N}_c, \label{con2P}
		\end{alignat}
	\end{subequations}
	where $E_2$ is a constant and can be obtained as follows
	
	\begin{equation}\label{E_2}
		\begin{split}
			E_2=& P^{(\mbox{\scriptsize hov})}
			\max\limits_{\substack{\forall n_i\in \mathcal{N}}}\Biggl\{
			\frac{\eta_{i}\mu_{ik}L}{\bar{\mathcal{R}}_{ik}(\bm{\eta},\bm{\mu},\bm{\psi})},   ~\mathcal{T}_{ik}(\bm{\mu},\textbf{f})\Biggr\}.
		\end{split}
	\end{equation}
	
	By introducing the auxiliary variables $\bm{\tau}\triangleq\{\tau_1, \cdots, \tau_{N_c}\}$ and $\varrho $, the optimization problem \textbf{P1-P} can be rewritten as 
	
	\begin{subequations}\label{P2_P}
		\begin{alignat}{2}
			\mbox{\textbf{P2-P}}~	&\!\min_{\substack{\textbf{P}_k},\bm{\tau}, \varrho}&&~~\varrho+\frac{1}{ P^{(\mbox{\scriptsize hov})}}\sum_{i=1}^{N_c}P_{ik} \tau_i, \label{OPP2P}\\
			&\mbox{s.t.}&&   P^{(\mbox{\scriptsize hov})} \mu_{ik} \frac{\sum\limits_{j=1}^{C}r_{ij}M_j}{\mathcal{R}_{ik}} = \tau_i, \forall n_i \in \mathcal{N}_c,  \label{con3P2} \\
			&& &  \tau_i \leq \varrho, \forall n_i \in \mathcal{N}_c, \label{con4P2}\\
			&& &  E_2 \leq \varrho, \\
			&& & \sum_{i=1}^{N_c} \mu_{ik}P_{ik}  \leq P_k,   \label{con1P2}\\
			&&     &  P_{ik} \geq 0 ~ \forall   n_i\in \mathcal{N}_c. \label{con2P2}
		\end{alignat}
	\end{subequations}
	The corresponding Lagrangian function can be written as	
	\begin{equation}
		\begin{split}
			&	\mathcal{L}(\textbf{P}_k,\bm{\tau}, \varrho, \bm{\nu})=  \varrho +\frac{1}{ P^{(\mbox{\scriptsize hov})}}\sum_{i=1}^{N_c}P_{ik} \tau_i  + \sum_{i=1}^{N_c}\bar{\nu}_i\left( \tau_i - \varrho\right)   \\
			&+ \sum_{i=1}^{N_c} \nu_i \left(P^{(\mbox{\scriptsize hov})}  \mu_{ik} \sum\limits_{j=1}^{C}r_{ij}M_j - \tau_i\mathcal{R}_{ik}\right)   + \bar{\bar{\nu}}( E_2 - \varrho),~~~~~
		\end{split}
	\end{equation}
	where  $\bm{\nu} \triangleq\{\nu_i, \bar{\nu}_i, \bar{\bar{\nu}}\} $       are the Lagrangian multipliers. The Lagrange dual function can be written as  \cite[Sec. 5.1.2]{boyd2004convex}
	
	\begin{equation}
		\begin{split}
			\mathcal{D}(\bm{\nu})=	\min_{\substack{\textbf{P}_k,\bm{\tau}, \varrho}} 	\mathcal{L}(\textbf{P}_k,\bm{\tau}, \varrho, \bm{\nu}).
		\end{split}
	\end{equation}
	Further, the dual problem can be expressed as 
	
	\begin{subequations}
		\begin{alignat}{2}
			&\!\max_{\substack{\bm{\nu}}}&& ~~    \mathcal{D}(\bm{\nu})\\
			&\mbox{s.t.}&&   \bar{\nu}_i, \bar{\bar{\nu}} \geq 0. 
		\end{alignat}
	\end{subequations}
	The optimum
	solution should satisfy the following conditions

	\begin{equation}\label{Lag2}
		\begin{split}
			&\pdv{\mathcal{L}(\textbf{P}_k, \bm{\tau}, \varrho, \bm{\nu})}{P_{ik}}= 
			\frac{	\tau_i^*}{P^{(\mbox{\scriptsize hov})}} - {\nu_i}^*\tau_i^*\mathcal{R}_{ik}' =0,\\
			& 	\pdv{\mathcal{L}(\textbf{P}_k, \bm{\tau}, \varrho, \bm{\nu})}{\tau_i}=  \frac{P_{ik}^*}{P^{(\mbox{\scriptsize hov})}}  - {\nu_i}^*\mathcal{R}_{ik} +  \bar{\nu}_i^*=0,\\
			& \pdv{\mathcal{L}(\textbf{P}_k, \tau_i, \varrho, \bm{\nu})}{{\nu}_i}=P^{(\mbox{\scriptsize hov})}   \sum\limits_{j=1}^{C}r_{ij}M_j - \tau_i^*\mathcal{R}_{ik}=0,\\
			&  P_{ik}^*\geq0,
		\end{split}
	\end{equation}
	where  	$\mathcal{R}_{ik}'\triangleq\pdv{\mathcal{R}_{ik}}{P_{ik}}=\frac{B|{\mathbf{h}^H_{ik}}{\mathbf{w}_{ik}}|^2}{P_{ik}^*|{\mathbf{h}^H_{ik}}{\mathbf{w}_{ik}}|^2+\omega_i}$, with $\omega_i=\sum\limits_{\substack{i'=1\\i'\ne i}}^{N}P_{i'k}^*|{\mathbf{h}^H_{ik}}{\mathbf{w}_{i'k}}|^2
	+ \sigma^2_i$.  By setting $\varrho =\tau_i $ and  $\bar{\nu}_i^*=0 $,  the power allocation can be  obtained using \eqref{Lag2} as follows

\begin{equation}\label{POptimum1}
	P_{ik}^* = \begin{cases}     \frac{P^{(\mbox{\scriptsize hov})}  \omega_i\sum\limits_{j=1}^{C}r_{ij}M_j}{\varrho^* B|{\mathbf{h}^H_{ik}}{\mathbf{w}_{ik}}|^2 - P^{(\mbox{\scriptsize hov})} \sum\limits_{j=1}^{C}r_{ij}M_j}, &\mbox{if}~ \mu_{ik} =1,   \\ 
		0, & \mbox{otherwise}. \end{cases}
\end{equation}
Finally, it can be noticed that $ 	P_{ik} $ is monotonically decreasing with respect to $\varrho$ and the feasibility range of $\varrho$ is $\varrho^* \in [\varrho_{\min}, \varrho_{\max}]$. The lower limit can be obtained as 
\begin{equation}
	\varrho_{\min}=\max \{E_2,\frac{P^{(\mbox{\scriptsize hov})}}{B|{\mathbf{h}^H_{ik}}{\mathbf{w}_{ik}}|^2} \sum\limits_{j=1}^{C}r_{ij}M_j \forall n_i\in \mathcal{N}_c\}.
\end{equation}  
The upper limit can be obtained  as follows. Let us define $\varrho_P$ as the value that satisfies \eqref{con3P2} and \eqref{con4P2}, which  corresponds to  equally allocate the transmission power and can be expressed  as follows
\begin{equation}\label{varrhoMax}
	\varrho_{P} = 	\max_{\substack{\forall n_i\in \mathcal{N}_c}} \left\{P^{(\mbox{\scriptsize hov})}  \frac{\sum\limits_{j=1}^{C}r_{ij}M_j}{{R}_{ik}} \right\},
\end{equation}
where ${R}_{ik}$ is the data rate of equally allocate the  transmission power (i.e., $P_{{i}k}=P_k/N_c$).
The upper limit can be obtained as $\varrho_{\max}=\max\{\varrho_{P},E_2\}$.
Algorithm \ref{PAlgorithm} illustrates the   procedure of solving Problem \textbf{P1-P}.

\begin{algorithm}
	\small 
	\caption{Transmission power allocation.}\label{PAlgorithm}
	\begin{algorithmic}[1]
		\State    \textbf{Input:} $\bm{\eta},\bm{\mu}, \bm{\psi},   \textbf{f}_k$,    $P_k$, and  $\epsilon$;
		\State    \textbf{Initialize:}  $\varrho_{\min}=E_2$; $P_{ik}=P_k/N_c$; Obtain $\varrho_{P}$ using \eqref{varrhoMax}; $\varrho_{\max}=\max\{\varrho_{P
		},E_2\}$; Set  $\varrho^*\leftarrow\varrho_{\max} $; 
		\State Obtain $		P_{ik}^*, ~\forall   n_i\in \mathcal{N}_c$ using \eqref{POptimum1};
		\State \textbf{While} $|\varrho_{\max}-\varrho_{\min}| >\epsilon$    \textbf{do}:
		\State $~~$ $\varrho = \frac{\varrho_{\max} +\varrho_{\min}}{2}$;    	
		\State $~~$  Obtain $		P_{ik}, ~\forall   n_i\in \mathcal{N}_c$ using \eqref{POptimum1};
		\State $~~$ \textbf{If}    $ \sum_{i=1}^{N_c} P_{ik}  \leq P_k $ \textbf{and} $	P_{ik} \geq 0~ \forall   n_i\in \mathcal{N}_c$ \textbf{do}:
		\State $~~~~$   $\varrho_{\max}\leftarrow\varrho $;
		\State $~~~~$   $	P_{ik}^*  \leftarrow	P_{ik}$;
		\State $~~$ \textbf{Else do} 
		\State $~~~~$   $\varrho_{\min}\leftarrow\varrho $;
		\State $~~$ \textbf{End If}
		\State \textbf{Return}  $P_{ik}^*, ~\forall   n_i\in \mathcal{N}_c$.
		\normalsize
	\end{algorithmic}
\end{algorithm}

\subsection{Optimizing the Computing Resources}
For a given  $\bm{\eta},\bm{\mu}, \bm{\psi},   \textbf{P}$, and $K$,    the edge computing resource of UAV $u_k$ can be obtain by solving the following optimization problem
\begin{subequations}\label{P1_F}
	\begin{alignat}{2}
		\mbox{\textbf{P1-f}}~	&\!\min_{\substack{\textbf{f}_k}}&&\max\limits_{\substack{\forall n_i\in \mathcal{N}}}\!\!\left\{\! P^{(\mbox{\scriptsize hov})}   \mu_{\tilde{\tilde{i}}k}\left(\frac{\ell_{\tilde{\tilde{i}}}}{\bar{\mathcal{R}}_{\tilde{\tilde{i}}k}} + \frac{\ell_{\tilde{\tilde{i}}}\varsigma_{\tilde{\tilde{i}}}}{f_{i\tilde{\tilde{i}}k}}\right), E_1\! \right\} +\kappa \sum_{i=1}^{N_m}\ell_{\tilde{\tilde{i}}}\varsigma_{\tilde{\tilde{i}}} f_{{\tilde{\tilde{i}}}k}^2\mu_{\tilde{\tilde{i}}k}, \\
		&\mbox{s.t.}&& \sum_{i=1}^{N_m}\mu_{\tilde{\tilde{i}}k} f_{{\tilde{\tilde{i}}}k}  \leq F_k,   \label{con1F}\\
		&&     &  f_{{\tilde{\tilde{i}}}k} \geq 0 ~ \forall   n_i\in \mathcal{N}_m, \label{con2F}
	\end{alignat}
\end{subequations}
where $E_1$ is a constant and can be obtained as follows
\begin{equation}\label{E_1}
	\begin{split}
		E_1=& P^{(\mbox{\scriptsize hov})}
		\max\limits_{\substack{\forall n_i\in \mathcal{N}}}\Biggl\{  \frac{\mu_{ik}\sum\limits_{j=1}^{C}r_{ij}M_j}{\mathcal{R}_{ik}(\bm{\mu},\bm{\psi}, \textbf{P})},  
	   \frac{\eta_{i}\mu_{ik}L}{\bar{\mathcal{R}}_{ik}(\bm{\eta},\bm{\mu},\bm{\psi})}\Biggr\}.
	\end{split}
\end{equation}
By introducing the auxiliary variable $\lambda$, the optimization problem \textbf{P1-f} can be rewritten as follows 
  
\begin{subequations}\label{P2_F}
	\begin{alignat}{2}
		\mbox{\textbf{P2-f}}~	&\!\min_{\substack{\textbf{f}_k, \lambda}}&& ~~ \lambda +\kappa \sum_{i=1}^{N_m}\ell_{\tilde{\tilde{i}}}\varsigma_{\tilde{\tilde{i}}} f_{{\tilde{\tilde{i}}}k}^2\mu_{\tilde{\tilde{i}}k}, \label{OPP2F}\\
		&\mbox{s.t.}&& P^{(\mbox{\scriptsize hov})} \mu_{\tilde{\tilde{i}}k}\left(\frac{\ell_{\tilde{\tilde{i}}}}{\bar{\mathcal{R}}_{{\tilde{\tilde{i}}}k}} + \frac{\ell_{\tilde{\tilde{i}}}\varsigma_{\tilde{\tilde{i}}}}{f_{{\tilde{\tilde{i}}}k}}\right) \leq \lambda, \forall n_i\in \mathcal{N}_m  \label{con3F2}\\
				&& & E_1  \leq \lambda,   \label{con0F2}\\
		&& & \sum_{i=1}^{N_m} \mu_{\tilde{\tilde{i}}k}f_{{\tilde{\tilde{i}}}k}  \leq F_k,   \label{con1F2}\\
		&&     &  f_{{\tilde{\tilde{i}}}k} \geq 0 ~ \forall   n_i\in \mathcal{N}_m. \label{con2F2}
	\end{alignat}
\end{subequations}
The corresponding Lagrangian function can be written as

\begin{equation}
	\begin{split}
		\mathcal{L}(\textbf{f}_k, \lambda,\bm{\chi})&=  \lambda +\kappa \sum_{i=1}^{N_m}\ell_{\tilde{\tilde{i}}}\varsigma_{\tilde{\tilde{i}}} f_{{\tilde{\tilde{i}}}k}^2 \mu_{\tilde{\tilde{i}}k}  
		+\chi_1\left(E_1-\lambda\right)+ \!\sum_{i=1}^{N_m}\bar{\chi}_i\! \left(P^{(\mbox{\scriptsize hov})}\mu_{\tilde{\tilde{i}}k}\!\left(\!\frac{\ell_{\tilde{\tilde{i}}}}{\bar{\mathcal{R}}_{\tilde{\tilde{i}}k}}\! +\! \frac{\ell_{\tilde{\tilde{i}}}\varsigma_{\tilde{\tilde{i}}}}{f_{{\tilde{\tilde{i}}}k}}\!\right) -\! \lambda\!\right),
	\end{split}
\end{equation}
 where $ \bm{\chi} \triangleq \{\chi_1,  \bar{\chi}_1, \cdots, \bar{\chi}_{N_m}\} $ represents the Lagrangian multipliers. The Lagrange dual function can be expressed as  \cite[Sec. 5.1.2]{boyd2004convex}
   
   \begin{equation}
   	\begin{split}
   \mathcal{D}(\bm{\chi})=	\min_{\substack{\textbf{f}_k, \lambda}} \mathcal{L}(\textbf{f}_k, \lambda,\bm{\chi}).
   	\end{split}
   \end{equation}
 The dual problem can be formulated as 

\begin{subequations}
	\begin{alignat}{2}
	&\!\max_{\substack{\bm{\chi}}}&& ~~    \mathcal{D}(\bm{\chi})\\
		&\mbox{s.t.}&& \chi_1,  \bar{\chi}_1, \cdots, \bar{\chi}_{N_m} \geq 0. 
	\end{alignat}
\end{subequations} 
 Given that $\mu_{\tilde{\tilde{i}}k}$ is a binary variable and there is no need to allocate computing resources at $u_k$ for the users that are not associated with $u_k$, it can be deduced that  $f_{\tilde{\tilde{i}}k}^*=0$ for all users with $\mu_{\tilde{\tilde{i}}k}=0$.   For a fixed Lagrangian multipliers and by applying the Karush-Kuhn-Tucker (KKT)
conditions and forcing \eqref{con3F2} into equality, a solution for $\mbox{\textbf{P2-f}}$  should satisfy the following conditions 

 \begin{equation}\label{Lag1}
 	\begin{split}
 		&\pdv{\mathcal{L}(\textbf{f}_k, \lambda,\bm{\chi})}{f_{{\tilde{\tilde{i}}}k}}= 2\kappa  \ell_{\tilde{\tilde{i}}}\varsigma_{\tilde{\tilde{i}}} f_{\tilde{\tilde{i}}k}^*  - \bar{\chi}_i^*P^{(\mbox{\scriptsize hov})} \frac{\ell_{\tilde{\tilde{i}}}\varsigma_{\tilde{\tilde{i}}}}{{f^*_{\tilde{\tilde{i}}k}}^2} =0,\\
 		& 	\pdv{\mathcal{L}(\textbf{f}_k, \lambda,\chi)}{\bar{\chi}_i}=  P^{(\mbox{\scriptsize hov})}\!\left(\!\frac{\ell_{\tilde{\tilde{i}}}}{\bar{\mathcal{R}}_{\tilde{\tilde{i}}k}}\! +\! \frac{\ell_{\tilde{\tilde{i}}}\varsigma_{\tilde{\tilde{i}}}}{f_{{\tilde{\tilde{i}}}k}^*}\!\right) -\! \lambda^*=0,\\
 		& f_{\tilde{\tilde{i}}k}^* \geq 0.
 	\end{split}
 \end{equation}
After simple manipulations and keeping in mind that $f_{{\tilde{\tilde{i}}}k}^*$ is a positive value, a solution for $\mbox{\textbf{P2-f}}$  can be obtained as 

\begin{equation}\label{FOptimum1}
	f_{{\tilde{\tilde{i}}}k}^* = \begin{cases} \frac{\ell_{\tilde{\tilde{i}}}\varsigma_{\tilde{\tilde{i}}}P^{(\mbox{\scriptsize hov})}}{\lambda^*-P^{(\mbox{\scriptsize hov})}\frac{\ell_i}{\bar{\mathcal{R}}_{ik}}}, &\mbox{if}~ \mu_{\tilde{\tilde{i}}k} =1,   \\ 
		0, & \mbox{otherwise}. \end{cases}
\end{equation}
Finally, it can be noticed that $ f_{{\tilde{\tilde{i}}}k} $ is monotonically decreasing with respect to $\lambda$ and the feasibility range of $\lambda$ is $\lambda^* \in [\lambda_{\min}, \lambda_{\max}]$. The lower limit can be obtained as $\lambda_{\min}=\max\{E_1,P^{(\mbox{\scriptsize hov})}\frac{\ell_i}{\bar{\mathcal{R}}_{ik}} \forall n_i \in \mathcal{N}_m\}$.  The upper limit can be obtained  as follows. Let us define $\lambda_F$ as the value that satisfies \eqref{con3F2}, which corresponds to     equally allocate the available computing resources and can be expressed  as follows
\begin{equation}\label{LambdaMax}
\lambda_{F} = 	\max_{\substack{\forall n_i\in \mathcal{N}_m}} \left\{P^{(\mbox{\scriptsize hov})} \left(\frac{\ell_{\tilde{\tilde{i}}}}{\bar{\mathcal{R}}_{{\tilde{\tilde{i}}}k}} + \frac{\ell_{\tilde{\tilde{i}}}\varsigma_{\tilde{\tilde{i}}}}{F_k/N_m}\right) \right\}.
\end{equation}
 The upper limit can be obtained as $\lambda_{\max}=\max\{\lambda_{F},E_1\}$.
  Algorithm \ref{FAlgorithm} illustrates the   procedure of solving Problem \textbf{P1-f}.

\begin{algorithm}
	\small 
	\caption{Optimizing the computing resources.}\label{FAlgorithm}
	\begin{algorithmic}[1] 
		\State    \textbf{Input:} $\bm{\eta},\bm{\mu}, \bm{\psi},   \textbf{P}$,    $F_k$, and  $\epsilon$;
		\State    \textbf{Initialize:}  $\lambda_{\min}=E_1$;  Obtain $\lambda_{F}$ using \eqref{LambdaMax}; $\lambda_{\max}=\max\{\lambda_{F},E_1\}$; Set  $\lambda^*\leftarrow\lambda_{\max} $;
			\State Obtain $	f_{{\tilde{\tilde{i}}}k}^*, ~\forall   n_i\in \mathcal{N}_m$ using \eqref{FOptimum1};
     	\State \textbf{While} $|\lambda_{\max}-\lambda_{\min}| >\epsilon$    \textbf{do}:
     	\State $~~$ $\lambda = \frac{\lambda_{\max} +\lambda_{\min}}{2}$;    	
     	\State $~~$  Obtain $	f_{{\tilde{\tilde{i}}}k}, ~\forall   n_i\in \mathcal{N}_m$ using  \eqref{FOptimum1};
     	\State $~~$ \textbf{If}    $ \sum_{i=1}^{N_m} f_{{\tilde{\tilde{i}}}k}  \leq F_k $ \textbf{and} $	f_{{\tilde{\tilde{i}}}k} \geq 0~ \forall   n_i\in \mathcal{N}_m$ \textbf{do}:
		\State $~~~~$   $\lambda_{\max}\leftarrow\lambda $;
		\State $~~~~$   $	f_{{\tilde{\tilde{i}}}k}^*  \leftarrow	f_{{\tilde{\tilde{i}}}k}$;
		\State $~~$ \textbf{Else do} 
		\State $~~~~$   $\lambda_{\min}\leftarrow\lambda $;
		\State $~~$ \textbf{End If}\color{black}
		\State \textbf{Return}  $f_{{\tilde{\tilde{i}}}k}^*, ~\forall   n_i\in \mathcal{N}_m$.
		\normalsize
	\end{algorithmic}
\end{algorithm}

\subsection{Graph-based Solution}
\begin{figure}[h!]
	\centering
	\includegraphics[width=0.6\textwidth]{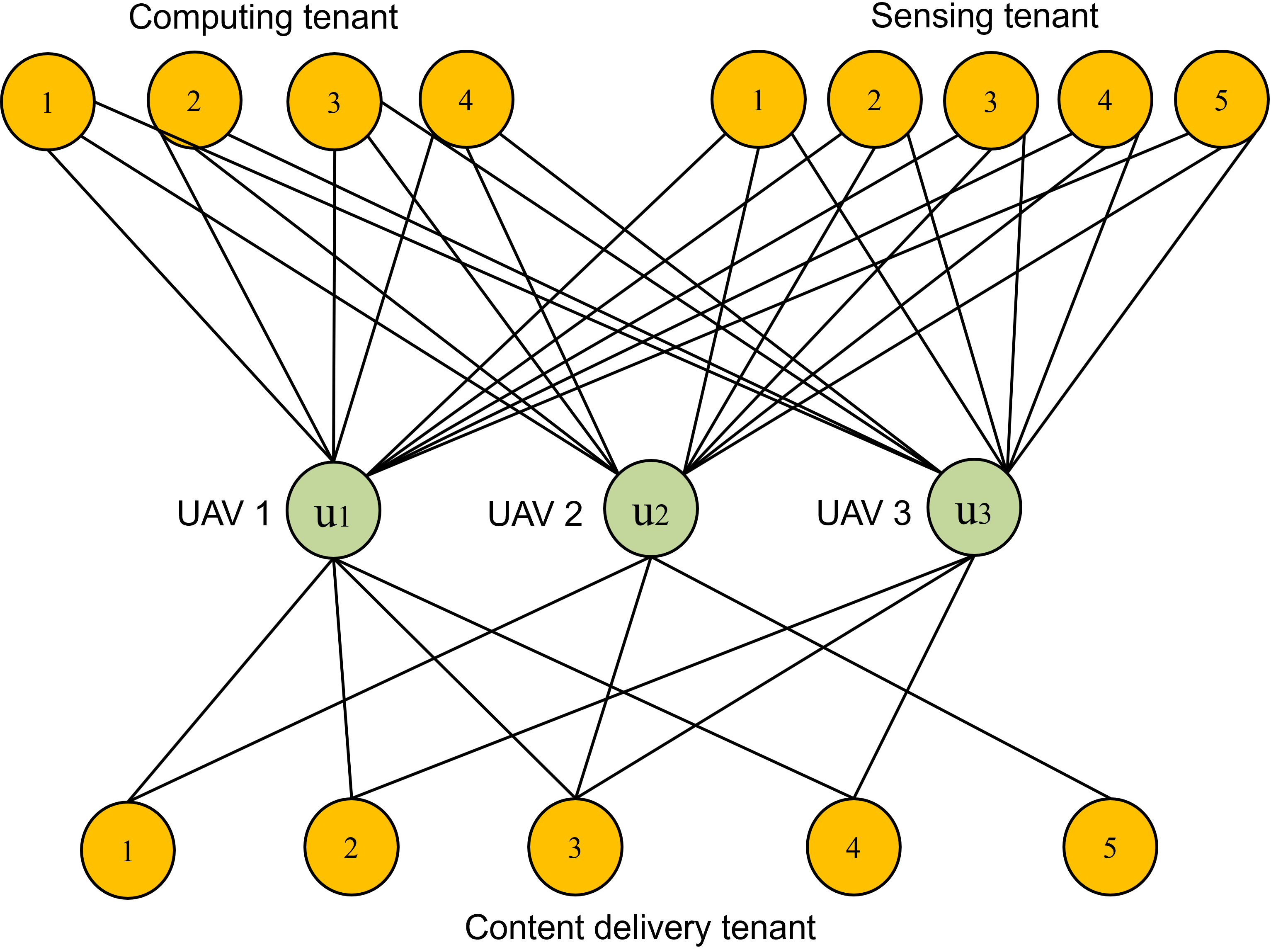}
	\caption{User-UAV    association graph of $3$ UAVs, $5$ sensing, $4$ computing, and $5$ content delivery users with the storage and demand in \eqref{demR}.}
	\label{graph1}
\end{figure}

To find the  optimized number of deployed UAVs,  the sensing users activation, and the user-UAV association,     a
representation of all feasible solutions should be first designed.  A quadripartite  graph model is proposed  as $\mathcal{G}\triangleq\{\mathcal{V}, \mathcal{E}, \mathcal{W}\}$, where $\mathcal{V}$ represents the vertices,  $ \mathcal{E}$ represents   the edges, and $\mathcal{W}$ is the weight of the edges.\\
\textbf{Vertex Set}:   The vertices set consists of four subsets  $\mathcal{V}=\{\mathcal{V}^c, \mathcal{V}^s, \mathcal{V}^m, \mathcal{V}^u\}  $.   The subset   $\mathcal{V}^c\triangleq\{v^c_{i}\}$  represents the   content delivery users such that $|\mathcal{V}^c|= N_c$,
  $\mathcal{V}^s\triangleq\{v^s_{i}\}$ represents the  sensing   users such that $|\mathcal{V}^s|= N_s$,  and $\mathcal{V}^m\triangleq\{v^m_{i}\}$  represents the    MEC users such that $|\mathcal{V}^m|= N_m$. 
The subset   $\mathcal{V}^u\triangleq\{v^u_{k}\}$    represents the UAVs    such that $|\mathcal{V}^u|=U$. \\
\textbf{Association Edges}: Each vertex in the subsets $\mathcal{V}^s$ and $\mathcal{V}^m$ is connected with all the vertices of $\mathcal{V}^u$ as a sensing user or MEC user can be served by any UAV.    An edge connects a vertex   $v_i^c \in \mathcal{V}^c$  and $v_k^u \in \mathcal{V}^u$ if $r_{ij}s_{jk}=1$, as the content delivery user can be served only by the UAVs that store its required content. It is worth noting that the total number of vertices in the graph $\mathcal{G}$ is $N+U$, and keeping in mind that $\mathcal{G}$ is quadripartite and no edge is needed within the users' vertices, the number of edges is upper-bounded by $NU$. This makes the graph size manageable for a reasonably large number of users and UAVs. 
Figure \ref{graph1} illustrates an example of the graph model with three UAVs,      five sensing users, five content delivery users, and four MEC users. The UAVs' stored contents   and the users' demands are as follows

\begin{equation}\label{demR}
	\mathbf{R}\!=\!\begin{blockarray}{cccccc}
	n_1 & n_2 & n_3 & n_4 & n_5  \\
	\begin{block}{[ccccc]c}
		0 & 0 & 1 & 0 & 0 & \!\!\! c_1\\
		0 & 0 & 0 & 0 & 1 & \!\!\! c_2 \\
		1 & 0 & 0 & 0 & 0 & \!\!\! c_3 \\
		0 & 1 & 0 & 1 & 0 & \!\!\! c_4 \\
	\end{block}
\end{blockarray}\!;
	\mathbf{S}\!=\!\begin{blockarray}{cccc}
		u_1 & u_2 & u_3  \\
		\begin{block}{[ccc]c}
			1 & 1 & 1 & \!\!\! c_1\\
			0 & 1 & 0 & \!\!\! c_2 \\
			1 & 1 & 0 & \!\!\! c_3 \\
			1 & 0 & 1 & \!\!\! c_4 \\
		\end{block}
	\end{blockarray}\!.
\end{equation}

\textbf{Edges' Weight}:  The weight of each edge should reflect the energy consumption of associating the corresponding user and UAV. The weights  of the graph     $\mathcal{W}=\left[w_{ik} \right]_{\footnotesize{ N\times U}}$  are updated as the users are associated,  such that  the weight of each edge $w_{ik}(\hat{\iota})$ at the $ \hat{\iota}$-th step is assigned as follows

\begin{equation}\label{Weight}
w_{ik}(\hat{\iota}) = \begin{cases} \frac{\Delta(\hat{\iota})}{ \frac{P^{(\mbox{\scriptsize mov})}d_{ik}(\hat{\iota})}{V}  + \frac{P^{(\mbox{\scriptsize hov})}L}{\bar{\mathcal{R}}_{ik}(\hat{\iota})}}, &\mbox{if} ~n_i \in \mathcal{N}_s,   \\ 
	\frac{1}{ \frac{P^{(\mbox{\scriptsize mov})}d_{ik}(\hat{\iota})}{V}  + P^{(\mbox{\scriptsize hov})}\left(\frac{\ell_i}{\bar{\mathcal{R}}_{\tilde{\tilde{i}}k}(\hat{\iota})} + \frac{\ell_i\varsigma_i}{f_{{\tilde{\tilde{i}}}k}}\right)}, &\mbox{if}~n_i \in \mathcal{N}_m,  \\ 
	\frac{\sum_{j=1}^{C}r_{ij}s_{jk}}{ \frac{P^{(\mbox{\scriptsize mov})}d_{ik}(\hat{\iota})}{V}  + \frac{(P^{(\mbox{\scriptsize hov})} +P_{ik})\sum_{j=1}^{C}r_{ij}s_{jk}M_j}{{\mathcal{R}}_{ik}(\hat{\iota})}},\!\!\!\! &\mbox{if} ~n_i \in \mathcal{N}_c, \end{cases} 
\end{equation} 
where $\Delta(\hat{\iota})=1$ if $\mathcal{H}(\hat{\iota}) < I$, with $\mathcal{H}(\hat{\iota})$ as the information gathered by the already associated sensing devices at the  $\hat{\iota}$-th step; otherwise $\Delta(\hat{\iota})=0$.  It is worth noting that the denominators in the first, second, and third lines of \eqref{Weight} represent the energy consumption of associating the UAV $u_k$ with the sensing user, MEC user, and content delivery user $n_i$, respectively. Consequently, an edge with the highest weight represents the lowest energy consumption of a user-UAV association.

The optimum altitude of the deployed UAV $ z_k $  depends on the required effective coverage area, with a circle whose  center is $(x_k, y_k)$ and whose radius is the distance between the center and the farthest associated user. Let $\vartheta_{ik} =\sqrt{(\bar{x}_i-x_k)^2 +(\bar{y}_i-y_k)^2}$ be the distance between user $n_i$ and the $x$-$y$ location of $u_k$, the optimum altitude can be
 expressed as $z_k= \hat{\vartheta}_{ik} \tan(\theta_{\mbox{\scriptsize opt}})$, where $ \hat{\vartheta}_{ik}= \max\limits_{1\leq i\leq N} \{\mu_{ik}\vartheta_{ik}\} $  and $ \theta_{\mbox{\scriptsize opt}} = 75.52^{\circ}$,   $54.62^{\circ}$, $42.44^{\circ}$, and $  20.34^{\circ} $ for the high-rise
urban, dense urban, urban,  and suburban environments, respectively \cite{alzenad20173}.
Keeping in mind the attitude limits in \eqref{con5}, the altitude of  UAV $u_k$ is obtained as follows

\begin{equation}\label{h_opt}
	z_k = \begin{cases} z^{\max}, & \mbox{if }  \hat{\vartheta}_{ik} \tan(\theta_{\mbox{\scriptsize opt}}) \geq z^{\max}, \\ 
			z^{\min}, &  \mbox{if }  \hat{\vartheta}_{ik} \tan(\theta_{\mbox{\scriptsize opt}}) \leq z^{\min}, \\ 
			\hat{\vartheta}_{ik} \tan(\theta_{\mbox{\scriptsize opt}}), & \mbox{otherwise}.\\ \end{cases}
\end{equation}

Algorithm \ref{euclid1} illustrates the proposed solution approach to solve \textbf{P1}.  The algorithm starts with constructing the  user-UAV association graph. It is worth noting that according to \eqref{Weight}, the initial weights equal zero for the edges between the content delivery users and the UAVs that do  not store the corresponding required contents (i.e., $\sum_{j=1}^{C}r_{ij}s_{jk}=0$ for each of those edges).  The algorithms iterates $Q$ iterations. At each iteration, the  algorithm calculates the weights using \eqref{Weight} based on the location of the UAVs and the equally allocated computing resource and   transmit power of the UAVs  (lines 6-7).  The edge with the maximum weight    $w_{\hat{i}\hat{k}}(\hat{\iota})$  is selected (line 11), user $n_{\hat{i}}$ is associated with UAV $u_{\hat{k}}$, and the   deployment  location of $u_{\hat{k}}$ is updated to be closer to all   associated users (lines 11-13).  If the associated user is a sensing user, it is set  as an active user and the amount of the gathered data is updated (lines 15-17). The corresponding weights of the selected users are set to zero and all the remaining  nonzero weights are updates based on the new location of the UAV (lines 18-19).  The algorithm continues while there exists an edge with non-zero weight. Once all the users are associated, the transmit power and computing resources of each deployed UAV are obtained using Algorithm $1$ and Algorithm $2$, respectively (lines 22-23). The algorithm  examines the value of the objective function at each iteration and returns the best solution (lines 24-29). { It is worth mentioning that once the solution is obtained, the SDN/NFV controller informs each of the deployed UAVs of their deployment location, the user-UAV association, and the transmit power and computation resource allocation decisions.}

{\color{black}

\begin{algorithm} 
	\caption{\color{black}A heuristic algorithm for energy-efficient UAV-enabled network slicing. }\label{euclid1}
	\begin{algorithmic}[1]
		\State \textbf{Input:} $	\mathbf{R}$,	$\mathbf{S}$, $ I $,   ${\bar{\psi}}_i, \forall n_i \in \mathcal{N}$,    $\mathcal{N}$, ${\psi}_k(0)$, $P_k$, $F_k$, $\forall u_k \in \mathcal{U}$;
		\State Construct the user-UAV association graph; 
		\State Set $O_{\min} \leftarrow \infty$;
		\State  \textbf{For} $q=1$  \textbf{to}  $Q$  \textbf{do}
		\State Set $ \bm{\mu}\leftarrow\left[0 \right]_{\footnotesize{ N\times U}}$ and $ \bm{\eta}\leftarrow\left[0 \right]_{\footnotesize{ N_s\times 1}}$; $ \hat{\iota}=0$;		$\mathcal{H}(0)=0$;
				\State Calculate
		 $d_{ik}(0)=\sqrt{\left(x_k-\bar{x}_i\right)^2+\left(y_k-\bar{y}_k\right)^2} $;   
		 	\State Set  $f_{{\tilde{\tilde{i}}}k}=F_k/N_m$;  $P_{ik}=P_k/\sum_{i=1}^{N_c}\sum_{j=1}^Cs_{jk}r_{ij}$;
		\State Obtain $w_{ik}(0)$  $\forall n_i ~ \in~ \mathcal{N}, \forall u_k  ~ \in~ \mathcal{U}$, using \eqref{Weight};		
		\State \textbf{While} $\exists ~ w_{ik}(\hat{\iota})  > 0~\forall  i=1, \dots, N, k=1, \dots, U $   \textbf{do}		
		\State $~$ $\hat{\iota}=\hat{\iota}+1$;  $ d_{ik}(\hat{\iota}) \leftarrow d_{ik}(\hat{\iota}-1)$; $ w_{ik}(\hat{\iota}) \leftarrow w_{ik}(\hat{\iota}-1)$, $\forall n_i \in \mathcal{N}, \forall u_k \in\mathcal{U}$;		
		\State $~$  Select $ \hat{i}$  and $ \hat{k}$  such that   \begin{equation*}
			(\hat{i},\hat{k})=\arg \max\limits_{\substack{1\leq i\leq N\\ 1 \leq k \leq U}} \{w_{ik}(\hat{\iota})\}.
		\end{equation*}
		\State $~$ Set $\mu_{\hat{i}\hat{k}}=1$;   $ x_{\hat{k}}(\hat{\iota})=\frac{\sum_{i=1}^{N}\mu_{i{\hat{k}}}\bar{x}_i}{\sum_{i=1}^{N}\mu_{i{\hat{k}}}}, $ $ ~ $ $ y_{\hat{k}}(\hat{\iota})=\frac{\sum_{i=1}^{N}\mu_{i\hat{k}}\bar{y}_i}{\sum_{i=1}^{N}\mu_{i\hat{k}}} $;
		\State $~$  $d_{i{\hat{k}}}(\hat{\iota})=\sqrt{\left(x_{\hat{k}}(\hat{\iota})-\bar{x}_i\right)^2+\left(y_{\hat{k}}(\hat{\iota})-\bar{y}_k\right)^2} $ $\forall ~ n_i ~ \in \mathcal{N}$;
		\State $~$	Obtain $ z_{\hat{k}} $   using \eqref{h_opt};
		\State $~$  \textbf{If}   $n_{\hat{i}} \in \mathcal{N}_s$ \textbf{do}
		\State $~~~$  Set $\eta_{\hat{i}}=1$; Calculate $\mathcal{H}(\hat{\iota})$ and $\Delta(\hat{\iota})$;
		\State $~$  \textbf{End If} 
		\State $~$  Set $w_{\hat{i}{k}}(\hat{\iota}) =0$ $\forall k=1, \cdots, U$;
		\State $~$  Update  $w_{i{k}}(\hat{\iota})$ if $w_{i\hat{k}}(\hat{\iota}) \neq 0$,  $\forall n_i \in \mathcal{N}, u_k \in \mathcal{U}$ using \eqref{Weight};
		\State   $ ~ $ \textbf{End While} 
			\State	$ ~ $ Set $K= |\mathcal{K}|$, where $\mathcal{K}$ is a set of deployed UAVs, $u_k \in \mathcal{K}$ if $\exists \mu_{ik}=1, \forall n_i \in \,\mathcal{N}$; 
			\State	$ ~~ $	Obtain $\textbf{P}_{{k}}$   $\forall u_k\in \mathcal{K}$  using Algorithm \ref{PAlgorithm};
 		\State	$ ~~ $	Obtain $\textbf{f}_{{k}}$     $\forall u_k\in \mathcal{K}$ using Algorithm \ref{FAlgorithm};
	    \State $ ~~ $	Calculate the  objective $O=\sum_{k=1}^{K}\mathcal{E}_k\left(\bm{\eta},\bm{\mu},\bm{\psi},\textbf{f},\textbf{P}\right)$;	 		 
	 \State $~~$  \textbf{If}   $ O < O_{\min} $ \textbf{do}	   
	 \State $~~~$   
	   $ O_{\min} =O$;  $K^*=K$;
	   \State $~~~$ $\bm{\mu}^* \leftarrow\bm{\mu}$,; $\bm{\eta}^*\leftarrow\bm{\eta}$; $\textbf{P}^*_k\leftarrow\textbf{P}_k$; $\textbf{f}^*_k\leftarrow\textbf{f}_k$; and ${\psi}^*_k\leftarrow{\psi}_k$;
	    	 \State $~~$  \textbf{End If}
	    	\State $~$  \textbf{End For}	 	    
		\State \textbf{Return}  $K^*$, $\bm{\mu}^* $, $\bm{\eta}^*$, $\textbf{P}^*_k$, $\textbf{f}^*_k$, and ${\psi}^*_k$.
%
%
%
		\normalsize
	\end{algorithmic}
\end{algorithm}

}

\subsection{Solution Computational Complexity Analysis}
	Constructing the user-UAV association graph   requires $\mathcal{O}(NU)$ operations,  while keeping in mind that $C\leq N$,   calculating the corresponding weights using \eqref{Weight} requires $\mathcal{O}(N^2U)$ operations. Obtaining the transmission power allocation  using Algorithm 1 requires $\mathcal{O}(\log(\frac{\varrho_{\max}-\varrho_{\min}}{\epsilon})N_c)$ operations. Further, optimizing the computing resources using Algorithm 2 requires $\mathcal{O}(\log(\frac{\lambda_{\max}-\lambda_{\min}}{\epsilon})N_m)$ operations.  Keeping in mind that $U \leq N$,   the computational complexity of solving \textbf{P1}  using Algorithm 3 can be expressed as  $\mathcal{O}(Q [ N^3 + \log(\frac{\varrho_{\max}-\varrho_{\min}}{\epsilon})N_c +\log(\frac{\lambda_{\max}-\lambda_{\min}}{\epsilon})N_m])$. It is worth noting that the computational complexity of Algorithm 3 is remarkably less than that of obtaining the   solution of \textbf{P1} using exhaustive search  (which is described in Section V), which can be expressed as   
$\mathcal{O}(U^N 2^{N_s}[ \log(\frac{\varrho_{\max}-\varrho_{\min}}{\epsilon})N_c +\log(\frac{\lambda_{\max}-\lambda_{\min}}{\epsilon})N_m])$.

 \section{Simulation Results}\label{SR}  
 This section introduces simulation results to evaluate the considered network slicing framework and the developed graph-based solution.  
In obtaining these results, it is assumed that the users are randomly placed within a square  
area of $ 1.2 \times 1.2 ~\mbox{km}^2 $. One-third   of the users are content delivery users,  one-third are sensing tenant users, and  one-third are MEC users.  
 The content storage capacity of a UAV is set as $75\%$  of the required contents and the content storage indicator matrix  $\mathbf{S}$ is obtained as a randomly-generated binary   matrix, such that $75\%$ of the elements in each column are ones  with the condition that the sum of each row is greater than or equal  $1$, to ensure that each content is stored at least in one UAV. 
	 The content demand matrix $\mathbf{R}$ is obtained as a randomly-generated binary  matrix with the condition that the sum of each row equals $1$, to ensure that each user   requests  one content.
  Each curve in the figures is obtained as the average of $1000$ Monte Carlo simulations.  Unless
otherwise mentioned, Table
\ref{TableResults} summarizes the    numerical values of the considered  system  parameters.

\begin{table}[h]
	\caption{Simulation Parameters.}
	\begin{center}
		\begin{adjustbox}{width=.6\textwidth}
			\begin{tabular}{|c|c||c|c||c|c|}
				\hline
				\textbf{Parameter}&\textbf{Value} & \textbf{Parameter}&\textbf{Value}   & \textbf{Parameter}&\textbf{Value}    \\ \hline
				$ N$ 	&   $ 27 $     &  $U $ 	& $ 5 $   &    $V$    & $12$ m/s \cite{8663615} \\ \hline  
				$ P_k$ 	&   $ 1 $ W  \cite{10547452} &  $ B_k $	&  $ 1 $ MHz  \cite{10547452}  &  $F_k$      &  $4$ GHz      \\  \hline  
				$ M_j$ 	&   $ 500\! $   Mbits  &  $\rho $ 	& $10^2 $   &     $L$   &  $ 1$ Mbits  \\ \hline  
				$ I$ 	&   $ 0.75 H $   \cite{9217066}    &  $F $ 	& $ 10 $ \cite{10547452}  & $\beta$       &   $ 2.2$ \cite{10547452} \\ \hline 
				$ \lambda_0$ 	&   $ -30 $ dB  \cite{10547452}   &  $v_0 $ 	& $ 4.03 $  \cite{8663615} & $\delta_0$       &   $ 0.6$ \cite{8663615} \\ \hline 	    	    
				$ \varsigma$ 	&   $  1.225 $ kg/m$^3$ \cite{8663615}   &  $\zeta $ 	& $ 0.05 $ \cite{8663615} &    $\xi$    & $0.503$ m$^2$ \cite{8663615}\\ \hline  
				$ \ell_i$ 	&   $  1 $ Mbits    &  $\varsigma_i $ 	& $ 700\times 10^8 $ cycle/bit  &    $\epsilon$    & $0.01$ \\ \hline  	    	   							             	    	   				 	 	 	 	          	    	   							             	    	   				 	 	 	 	          
			\end{tabular}
		\end{adjustbox}
	\end{center}
	\label{TableResults}
\end{table}

\begin{figure}[h!] 
	\centering
	\includegraphics[width=0.6\textwidth]{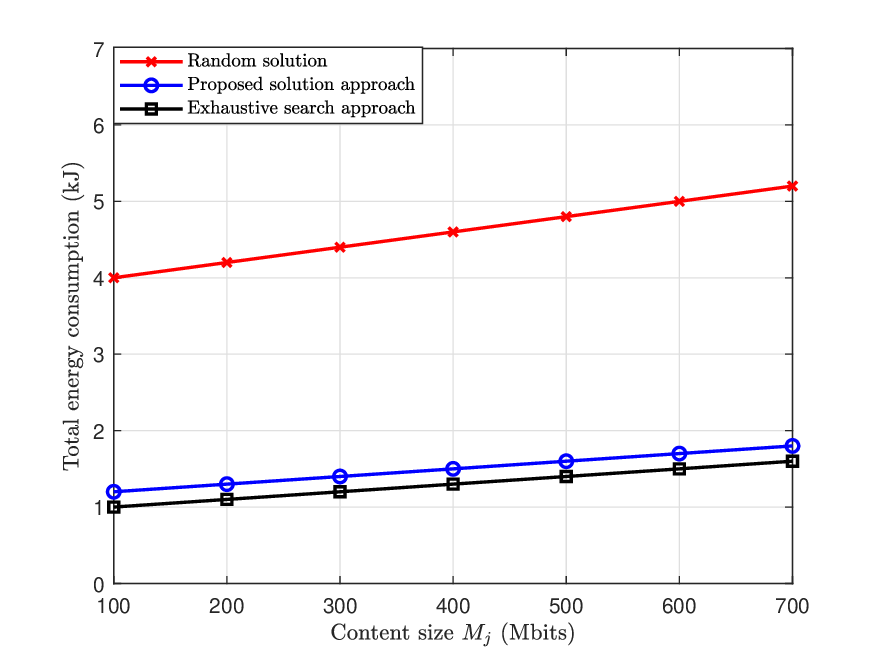}
	\caption{Total  energy consumption versus 	the content size with $N_c=3$, $N_s=2$, $N_m=2$, and  $ U=3 $.}
	\label{figMj}
\end{figure}

 Figure \ref{figMj} illustrates the total energy consumption versus the size of the contents $M_j$. Three solution approaches are illustrated: (I) \textit{Random solution:} in which all the available UAVs are deployed, each UAV is placed above a randomly selected user, each user is associated with the nearest UAV,  all the sensing users are active, and the transmit power and computing resources are equally allocated to the corresponding users. (II) \textit{Proposed solution approach:} which represents the performance of the graph-based solution in Algorithm \ref{euclid1}. (III) \textit{Exhaustive search approach:} in which all the possible user-UAV association and the sensing user activation are examined, while the transmission power and computing resources are obtained using Algorithm \ref{PAlgorithm} and Algorithm \ref{FAlgorithm}, respectively. It can be noticed that the proposed solution approach remarkably reduces the energy consumption in comparison with the random solution. On the other hand, the proposed  approach provides close to the optimum solution  obtained using the exhaustive search.

\begin{figure}[h!] 
	\centering
	\includegraphics[width=0.6\textwidth]{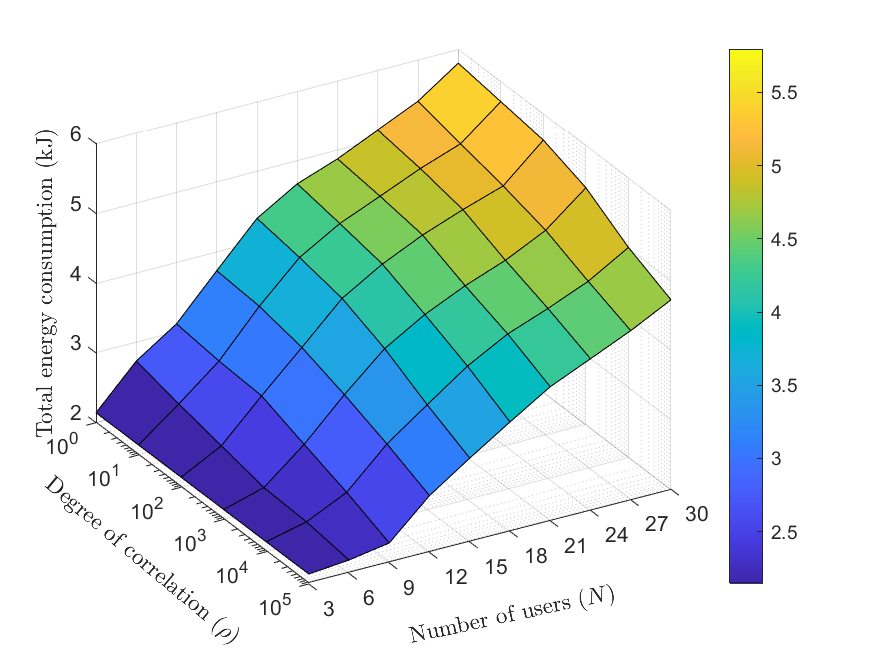}
	\caption{Total  energy consumption versus 	the   degree of correlation    $ \rho  $ and number of users $N$.}
	\label{fig3D}
\end{figure}

{Figure \ref{fig3D} provides  insight into the effect of the number of users and the degree of correlation    $ \rho  $ of the gathered data
	on the energy expenditure of the proposed algorithm. It can be noticed that the energy expenditure increases with the number of users. It can also be noticed that the energy expenditure decreases as the degree of correlation increases. This can be attributed to the fact that as the degree of correlation increases, the data of the users becomes more correlated, and thus, fewer sensing users will meet the sensing tenant requirement. Furthermore, this behaviour indicates that the proposed solution approach activates an adequate number of data-gathering users, and as the degree of correlation increases, fewer users are activated. }

\begin{figure}[h!] 
 	\centering
 	\includegraphics[width=0.6\textwidth]{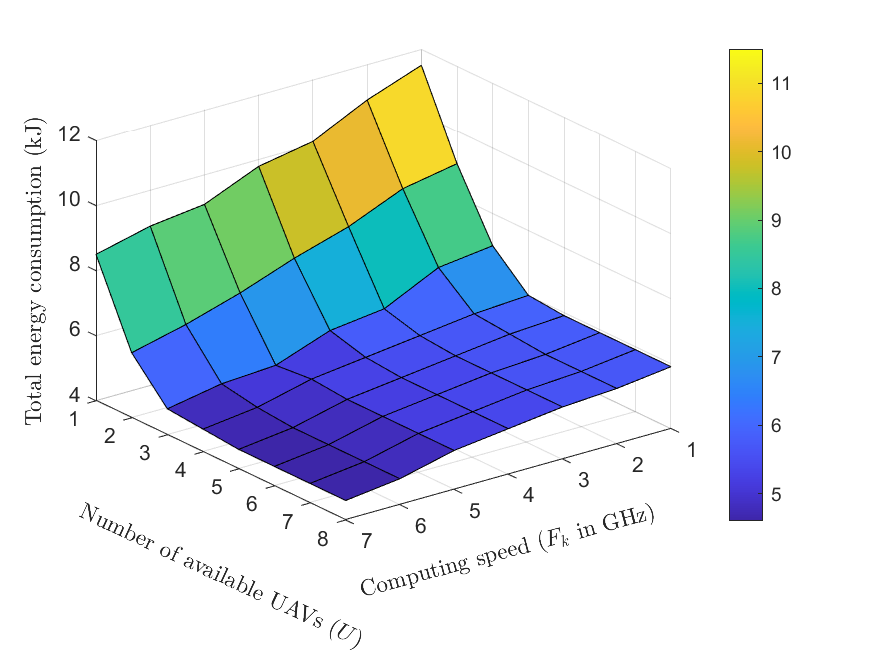}
 	\caption{Total  energy consumption versus 	the   degree of correlation    $ \rho  $ and number of users $N$.}
 	\label{figUFk}
 \end{figure}

To illustrate the effect of the number of the available UAVs and the computing speed, Fig. \ref{figUFk} illustrates the  energy expenditure performance obtained using the proposed algorithm. It can be seen that the energy expenditure decreases as more UAVs are made available, and the floor behaviour occurs as the number of available UAVs increases. Such behaviour indicates that the number of deployed UAVs is optimized, and the proposed solution approach avoids deploying redundant UAVs, which otherwise lead to higher energy consumption. It can also be noticed that the energy expenditure increases as the computing speed decreases, which can be attributed to the fact that a slower  computing speed at the UAV yields a longer  hovering time and more energy consumption.

\begin{figure}[h!] 
	\centering
	\includegraphics[width=0.6\textwidth]{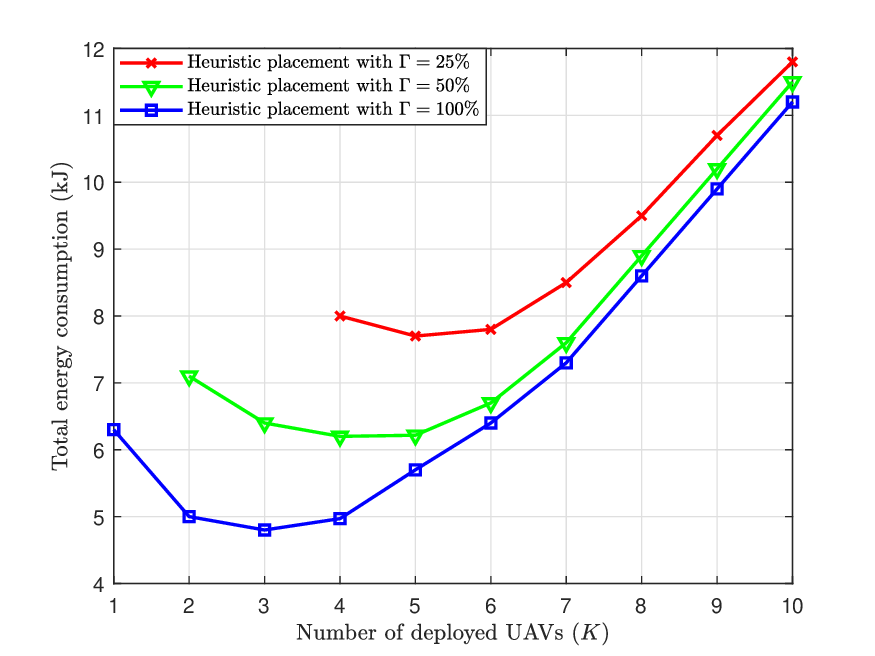}
	\caption{Total  energy consumption versus 	the number of deployed UAVs $K$ with different UAV storage capacity.}
	\label{figK}
\end{figure}

To study the effect of the number of deployed UAVs and their content storage capacity, Fig. \ref{figK} shows the energy consumption versus the number of deployed UAVs. This figure illustrates  a heuristic placement solution, in which the users are clustered using a $K$-means clustering technique, and $K$ UAVs are deployed  each serving a cluster of users. Three UAV content storage capacities are illustrated, such that $\Gamma=25\%$, $\Gamma=50\%$, and $\Gamma=100\%$ imply that each UAV stores  one-fourth, half, and  all the contents, respectively. It is worth mentioning that to guarantee that all  contents are available to users, at least four and two UAVs are deployed for $\Gamma=25\%$ and $\Gamma=50\%$, respectively. It can be noticed that as the storage capacity of the UAVs increases, less UAVs are required to be deployed and the corresponding energy consumption is less.  This is a result of the fact that increasing the UAV storage capacity enables it to serve more users. Consequently, fewer UAVs are deployed, and the energy consumed by the deployed UAVs is less than that of deploying a higher number of UAVs with redundant stored content.   It can be also noticed that deploying more UAVs increases the energy consumption as each UAV  consumes traveling energy and serves less users.

\begin{figure}[h!] 
	\centering
	\includegraphics[width=0.6\textwidth]{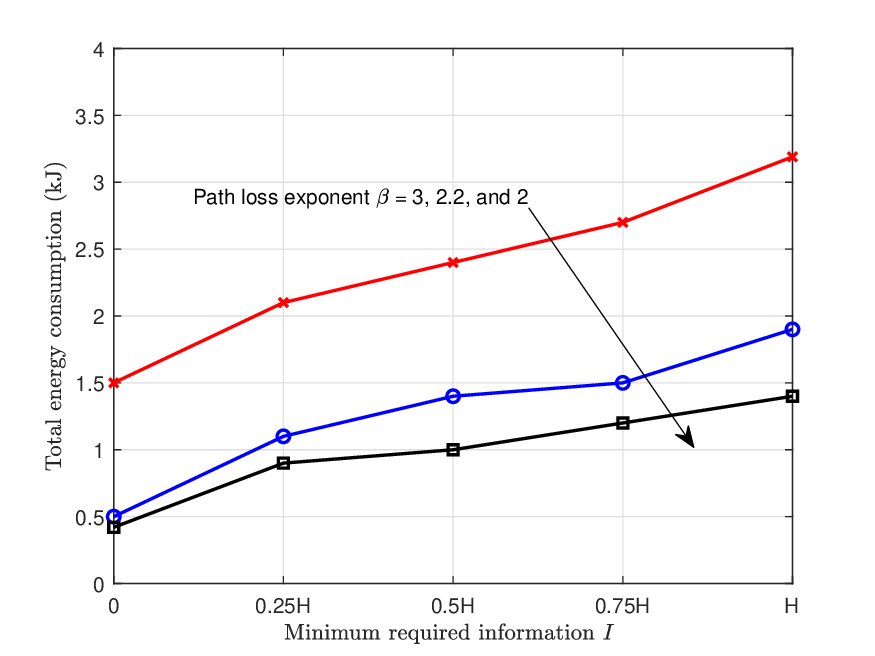}
	\caption{Total  energy consumption versus the	 minimum required information by the data gathering tenant $I$ for different values of the path loss exponent $\beta$.}
	\label{figI}
\end{figure}

The effect of the minimum required information by the data gathering tenant $I$ is illustrated in Fig. \ref{figI}. The considered range of $I$ lies between $0$ (which represents the extreme case, in which the tenant is not interested in gathering any information) and $H$ (which means the tenant is interested in gathering all the available information). It can be seen that the energy consumption increases with $I$, which can be attributed to the fact that as $I$ increases, more data-gathering users need to be activated and served by the UAVs. Furthermore, the figure depicts different values of the path loss exponent $\beta$. As the value of $\beta$ increases, more energy is consumed. This is a result of the fact that increasing the value of the path loss exponent $\beta$  reduces the achievable data rate, which leads to higher hovering time for the UAVs, and thus, increases the energy consumption. It is worth mentioning that changing the value of the Rician factor $F$ within a reasonable range does  not     noticeably affect   the total energy consumption.

\section{Conclusion}\label{CO}
This paper has introduced an energy-efficient network-slicing framework to deliver content and to gather information to/from the users as well as to provide MEC services.  The MEC tenant serves its users, the content delivery tenant mandates that each of its users receives the required content, and the sensing tenant needs to gather a sufficient amount of uncorrelated information.    An energy consumption minimization optimization has been formulated to meet the tenants' requirements. The spatial correlation among the sensing users has been considered to activate the necessary subset of  sensing users. A graph-based solution with the Lagrange  approach    has been developed. Simulation results have shown that the considered network-slicing framework   significantly reduces the total
energy consumption. This energy preservation is achieved by deploying a sufficient number of UAVs and optimally placing them, activating an adequate number of data-gathering users, efficiently associating the users with the deployed UAVs, and optimally allocating the transmit power and computing resources. Results also illustrated that the proposed solution provides performance close to the optimum one, with remarkably less computation complexity.  

\balance
\bibliographystyle{IEEEtran}
\bibliography{IEEEabrv,Refrences-library}

\begin{thebibliography}{10}
\providecommand{\url}[1]{#1}
\csname url@samestyle\endcsname
\providecommand{\newblock}{\relax}
\providecommand{\bibinfo}[2]{#2}
\providecommand{\BIBentrySTDinterwordspacing}{\spaceskip=0pt\relax}
\providecommand{\BIBentryALTinterwordstretchfactor}{4}
\providecommand{\BIBentryALTinterwordspacing}{\spaceskip=\fontdimen2\font plus
\BIBentryALTinterwordstretchfactor\fontdimen3\font minus
  \fontdimen4\font\relax}
\providecommand{\BIBforeignlanguage}[2]{{%
\expandafter\ifx\csname l@#1\endcsname\relax
\typeout{** WARNING: IEEEtran.bst: No hyphenation pattern has been}%
\typeout{** loaded for the language `#1'. Using the pattern for}%
\typeout{** the default language instead.}%
\else
\language=\csname l@#1\endcsname
\fi
#2}}
\providecommand{\BIBdecl}{\relax}
\BIBdecl

\bibitem{10551400}
W.~Rafique, J.~Rani~Barai, A.~O. Fapojuwo, and D.~Krishnamurthy, ``A survey on
  beyond {5G} network slicing for smart cities applications,'' \emph{IEEE
  Commun. Surv. Tutor.}, vol.~27, no.~1, pp. 595--628, Feb. 2025.

\bibitem{9780171}
S.~Zhao, G.~Qi, T.~He, J.~Chen, Z.~Liu, and K.~Wei, ``A survey of sparse mobile
  crowdsensing: Developments and opportunities,'' \emph{IEEE Open J. Comput.
  Soc.}, vol.~3, pp. 73--85, May 2022.

\bibitem{9217066}
A.~A. Al-habob, O.~A. Dobre, and H.~Vincent~Poor, ``Energy-efficient
  spatially-correlated data aggregation using unmanned aerial vehicles,'' in
  \emph{Proc. IEEE 31st Annual International Symposium on Personal, Indoor and
  Mobile Radio Communications}, 2020, pp. 1--6.

\bibitem{9409148}
A.~A. Al-Habob, O.~A. Dobre, and H.~V. Poor, ``Role assignment for
  spatially-correlated data aggregation using multi-sink internet of underwater
  things,'' \emph{IEEE Trans. Green Commun. Netw.}, vol.~5, no.~3, pp.
  1570--1579, Apr. 2021.

\bibitem{9606181}
------, ``Age- and correlation-aware information gathering,'' \emph{IEEE
  Wireless Commun. Lett.}, vol.~11, no.~2, pp. 273--277, Nov. 2022.

\bibitem{8169684}
A.~A. Al-Habob, Y.~N. Shnaiwer, S.~Sorour, N.~Aboutorab, and P.~Sadeghi,
  ``Multi-client file download time reduction from cloud/fog storage servers,''
  \emph{IEEE Trans. Mobile Comput.}, vol.~17, no.~8, pp. 1924--1937, Dec. 2018.

\bibitem{9174931}
A.~A. Al-Habob, O.~A. Dobre, S.~Muhaidat, and H.~Vincent~Poor,
  ``Energy-efficient data dissemination using a {UAV}: An ant colony
  approach,'' \emph{IEEE Wireless Commun. Lett.}, vol.~10, no.~1, pp. 16--20,
  Aug. 2021.

\bibitem{9865119}
A.~A. Al-Habob, O.~A. Dobre, S.~Muhaidat, and H.~V. Poor, ``Energy-efficient
  information placement and delivery using {UAVs},'' \emph{IEEE Internet Things
  J.}, vol.~10, no.~1, pp. 357--366, Jan. 2023.

\bibitem{10963879}
B.~Mao, Y.~Liu, Z.~Wei, H.~Guo, Y.~Xun, J.~Wang, J.~Liu, and N.~Kato, ``A
  blockchain-enabled cold start aggregation scheme for federated reinforcement
  learning-based task offloading in zero trust {LEO} satellite networks,''
  \emph{EEE J. Sel. Areas Commun.}, vol.~43, no.~6, pp. 2172--2182, Jun. 2025.

\bibitem{10547452}
A.~A. Al-habob, J.~Lin, O.~A. Dobre, and Y.~Jing, ``Min-max latency
  minimization for energy-constrained multi-{UAV} mobile edge computing,''
  \emph{IEEE Trans. Netw. Sci. Eng.}, vol.~11, no.~5, pp. 4577--4590, 2024.

\bibitem{9598918}
Z.~Xiao, L.~Zhu, Y.~Liu, P.~Yi, R.~Zhang, X.-G. Xia, and R.~Schober, ``A survey
  on millimeter-wave beamforming enabled {UAV} communications and networking,''
  \emph{IEEE Commun. Surv. Tutor.}, vol.~24, no.~1, pp. 557--610, First quarter
  2022.

\bibitem{10339670}
Y.~Kawamoto, M.~Takahashi, S.~Verma, N.~Kato, H.~Tsuji, and A.~Miura,
  ``Traffic-prediction-based dynamic resource control strategy in
  {HAPS}-mounted {MEC}-assisted satellite communication systems,'' \emph{IEEE
  Internet Things J.}, vol.~11, no.~8, pp. 13\,824--13\,836, Apr. 2024.

\bibitem{8698468}
C.~You and R.~Zhang, ``{3D} trajectory optimization in rician fading for
  {UAV}-enabled data harvesting,'' \emph{IEEE Trans. Wireless Commun.},
  vol.~18, no.~6, pp. 3192--3207, Apr. 2019.

\bibitem{10242032}
C.~De~Alwis, P.~Porambage, K.~Dev, T.~R. Gadekallu, and M.~Liyanage, ``A survey
  on network slicing security: Attacks, challenges, solutions and research
  directions,'' \emph{IEEE Commun. Surv. Tutor.}, vol.~26, no.~1, pp. 534--570,
  First quarter 2024.

\bibitem{10679214}
F.~Wei, G.~Feng, S.~Qin, Y.~Peng, and Y.~Liu, ``Hierarchical network slicing
  for {UAV}-assisted wireless networks with deployment optimization,''
  \emph{EEE J. Sel. Areas Commun.}, vol.~42, no.~12, pp. 3705--3718, Dec. 2024.

\bibitem{10847822}
F.~Wei, Y.~Wang, G.~Feng, and S.~Qin, ``Network slicing-enabled computation
  offloading in satellite-terrestrial edge computing networks: A bi-level game
  approach,'' \emph{IEEE Internet Things J.}, vol. Early Access, pp. 1--16,
  2025.

\bibitem{10988691}
N.~Xiong, W.~Zhong, Y.~Chen, D.~He, Y.~Li, L.~Chen, and W.~Liang, ``{MCDS}: An
  effective multi-{UAV} collaborative decision-making system in mobile-edge
  computing networks,'' \emph{IEEE Internet Things J.}, vol.~12, no.~15, pp.
  29\,354--29\,372, Aug. 2025.

\bibitem{11077695}
B.~Xiao, Z.~Yao, L.~Zhang, B.~Zhang, and C.~Li, ``An efficient online task
  offloading algorithm for bilevel {UAV}-enabled mobile edge computing,''
  \emph{IEEE Internet Things J.}, vol. Early Access, pp. 1--1, Jul. 2025.

\bibitem{11075956}
H.~H. Esmat, X.~Liu, B.~Lorenzo, and J.~Liu, ``Outage-aware multi-domain
  network slicing for satellite-airborne-terrestrial networks with multiple
  configurations,'' \emph{IEEE Trans. Wireless Commun.}, vol. Early Access, pp.
  1--1, Jul. 2025.

\bibitem{9127468}
G.~Faraci, C.~Grasso, and G.~Schembra, ``Design of a {5G} network slice
  extension with {MEC} {UAVs} managed with reinforcement learning,'' \emph{EEE
  J. Sel. Areas Commun.}, vol.~38, no.~10, pp. 2356--2371, Oct. 2020.

\bibitem{pattem2008impact}
S.~Pattem, B.~Krishnamachari, and R.~Govindan, ``The impact of spatial
  correlation on routing with compression in wireless sensor networks,''
  \emph{ACM Trans. Sensor Netw.}, vol.~4, no.~4, pp. 1--33, Aug. 2008.

\bibitem{8663615}
Y.~Zeng, J.~Xu, and R.~Zhang, ``Energy minimization for wireless communication
  with rotary-wing {UAV},'' \emph{IEEE Trans. Wireless Commun.}, vol.~18,
  no.~4, pp. 2329--2345, Mar. 2019.

\bibitem{8613833}
M.~B. {Ghorbel}, D.~{Rodríguez-Duarte}, H.~{Ghazzai}, M.~J. {Hossain}, and
  H.~{Menouar}, ``Joint position and travel path optimization for energy
  efficient wireless data gathering using unmanned aerial vehicles,''
  \emph{IEEE Trans. Veh. Technol.}, vol.~68, no.~3, pp. 2165--2175, Mar. 2019.

\bibitem{7542156}
Y.~Wang, M.~Sheng, X.~Wang, L.~Wang, and J.~Li, ``Mobile-edge computing:
  Partial computation offloading using dynamic voltage scaling,'' \emph{IEEE
  Trans. Commun.}, vol.~64, no.~10, pp. 4268--4282, Aug. 2016.

\bibitem{boyd2004convex}
S.~Boyd and L.~Vandenberghe, \emph{Convex Optimization}.\hskip 1em plus 0.5em
  minus 0.4em\relax Cambridge university press, 2004.

\bibitem{alzenad20173}
M.~Alzenad, A.~El-Keyi, F.~Lagum, and H.~Yanikomeroglu, ``{3-D placement of an
  unmanned aerial vehicle base station ({UAV-BS}) for energy-efficient maximal
  coverage},'' \emph{IEEE Wireless Commun. Lett.}, vol.~6, no.~4, pp. 434--437,
  Aug. 2017.

\end{thebibliography}

\end{document}